\documentclass{article}

\usepackage{arxiv}

\usepackage[utf8]{inputenc} 
\usepackage[T1]{fontenc}    
\usepackage{hyperref}       
\usepackage{url}            
\usepackage{booktabs}       
\usepackage{amsfonts}       
\usepackage{nicefrac}       
\usepackage{microtype}      
\usepackage{graphicx}
\usepackage{natbib}
\usepackage{doi}
\usepackage{subcaption}
\usepackage{amsmath}
\usepackage{cleveref}       
\usepackage{mathrsfs}
\usepackage[ruled,vlined]{algorithm2e}
\usepackage{bm}
\usepackage{multirow} 
\usepackage{arydshln}
\usepackage{dsfont}
\usepackage{tikz}
\usetikzlibrary{calc}
\usetikzlibrary{positioning, arrows.meta, fit}

\definecolor{lightgray}{RGB}{230,230,230}
\definecolor{lightblue}{RGB}{200,230,255}

\tikzset{
    latent/.style={circle, draw, fill=lightblue, minimum size=1cm},
    observed/.style={circle, draw, fill=lightgray, minimum size=1cm},
    param/.style={rectangle, draw, minimum size=0.9cm},
    arrow/.style={->, >=Stealth, thick},
    node style/.style={circle, draw, fill=white, minimum size=1.2cm, font=\bfseries}
}

\title{A Bayesian Joint Model for Multiple Point Processes with Application to Presence-Only Data}

\date{}

\newif\ifuniqueAffiliation
\uniqueAffiliationtrue

\ifuniqueAffiliation
\author{ 
\href{https://orcid.org/0009-0005-5852-8179}
{\includegraphics[scale=0.06]{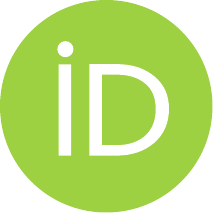}\hspace{1mm}Laura C. M. Teixeira} \thanks{Corresponding author: laurateixeiracm@gmail.com} \\
        Department of Statistics \\
	Federal University of Minas Gerais, Brazil\\
	\And
     \href{https://orcid.org/0000-0003-0697-4589}
	 {\includegraphics[scale=0.06]{orcid.pdf}\hspace{1mm}Dani Gamerman} \\
      Department of Statistical Methods \\
	 Federal University of Rio de Janeiro, Brazil \\
     \And
	 \href{https://orcid.org/0000-0002-3137-211X}
     {\includegraphics[scale=0.06]{orcid.pdf}\hspace{1mm}Vinicius Peripato} \\
     Division of Earth Observation and Geoinformatics\\
	 National Institute for Space Research, Brazil \\
     \And
	\href{https://orcid.org/0000-0002-8425-9479}
    {\includegraphics[scale=0.06]{orcid.pdf}\hspace{1mm}Carolina Levis} \\
        Graduate Program in Ecology \\
	Federal University of Santa Catarina, Brazil\\
}
\fi

\renewcommand{\shorttitle}{A Bayesian Joint Model for Multiple Point Processes}

\hypersetup{
pdftitle={A Bayesian Joint Model for Multiple Point Processes with Application to Presence-Only Data},
pdfsubject={stat.ME},
pdfauthor={Laura C. M. Teixeira, Dani Gamerman, Vinicius Peripato, Carolina Levis},
pdfkeywords={Multivariate Point Processes, Bayesian Networks, Poisson Process, Bayesian Inference, Joint Species Distribution Model, Amazon Forest},
}

\begin{document}
\maketitle

\begin{abstract}
Joint modeling of multiple point processes is relevant in applications where relationships among processes are of interest, such as in ecological and archaeological studies. Statistical inference becomes particularly challenging when multiple processes are analyzed jointly and the observed data correspond to presence-only patterns, which are subject to preferential sampling and partial observability. This paper proposes a Bayesian joint model for multiple point processes, with application to the presence-only setting. The dependence between processes is explicitly incorporated into the probabilistic specification of the model using Bayesian networks. 
Direct use of the likelihood leads to intractable likelihood functions.
Latent data processes are then introduced so that the augmented likelihood function becomes tractable and can be exactly evaluated. This formulation also enables direct inference on the number and the spatial distribution of unobserved occurrences of any of the point patterns. Inference is carried out using Markov chain Monte Carlo with blocked Gibbs sampling. Simulation studies demonstrate that the proposed inferential scheme is able to recover the true model parameters. 
The proposed model is applied to real presence-only data of archaeological sites and tree species from Amazonia, as part of the study of the effect that pre-Columbian Indigenous presence might have on the occurrences of relevant tree species.
The results are consistent with the findings reported in the literature. They also illustrate how the proposed model enables inference on the existence and on the magnitude of the relation between processes, in addition to their association with environmental covariates.
\end{abstract}


\section{Introduction}
\label{s:intro}

Understanding spatial point patterns is central to a wide range of problems in ecology and archaeology. A spatial point pattern is typically modeled as a realization of a point process, providing a probabilistic way to describe the distribution of events \citep{moller2003}. Many applications involve multiple events observed over a common spatial domain, giving rise to multitype point patterns. 
For example, in ecological systems, the spatial distribution of one species can influence that of another through mechanisms such as competition, facilitation, or hierarchical interactions \citep{Hougmander1999}.
In such settings, the joint modeling of multiple point patterns becomes essential for capturing interactions that cannot be inferred from independent analyses.

Several approaches have been proposed to jointly model multiple point patterns, particularly in ecological applications through joint species distribution models (JSDMs). These models typically incorporate dependence indirectly, for example through shared sampling effort surfaces or spatial random effects \citep{botella2021, escamilla2022}, or by sharing parameters related to detectability \citep{fithian2015, renner2019}. Although such formulations allow information to be pooled across species and may improve predictive performance, they do not explicitly represent interactions between processes. As a result, inference on the strength of relationships between species based on observational data remains limited.

This work proposes a Bayesian joint model for multiple point processes that addresses this limitation by explicitly modeling their dependence. 
This direct strategy to include dependence between processes is introduced directly by having point processes as additional covariates at the level of the intensity functions, enabling inference on the existence and the strength of such relationships. In this sense, the occurrence pattern of one process can influence the distribution of others, in addition to the effect of environmental covariates.

A modeling approach based on Poisson point processes is adopted, whose likelihood typically involves an intractable integral with no closed-form expression. To overcome this difficulty, a data augmentation strategy similar to that of \citet{goncalves2018} is used, resulting in a tractable likelihood and avoiding the need for numerical approximations commonly used in this context \citep{Warton2010,EquivalenceMaxent}. As a result, model parameters can be estimated by exact Bayesian inference, which has been shown to improve predictive performance and the quality of estimates \citep{moreira2022analysis, SilvaGamerman2026}.

Settings where data are collected opportunistically rather than through designed surveys are also considered, resulting in what is called presence-only data, in which only locations of observed events are available. Such data lack information on absences and on the underlying sampling effort, making statistical inference challenging. Despite these limitations, presence-only datasets are abundant and often represent the only available source of information.  

A key difficulty with presence-only data is that the sampling is typically biased. Records tend to concentrate in easily accessible regions or areas with higher observer activity, leading to preferential sampling. In this context, the mechanism that generates the observed data is stochastically related to the underlying point process. As a result, observed occurrences reflect a combination of the underlying process of interest and observation bias. 
The proposed model accounts for both components, enabling the analysis of multiple presence-only datasets while avoiding misleading inference.

Most existing approaches to modeling presence-only data focus on relating occurrence to environmental (abiotic) covariates, such as climate, topography, and soil characteristics. These models have been widely used in species distribution modeling (SDM), helping to characterize how environmental factors shape habitat suitability. However, they typically ignore biotic factors, such as interactions between species or between ecological and anthropogenic processes, despite evidence that these relations play a central role in shaping species distributions \citep{Kissling2012}. Ignoring then can also limit the interpretability and predictive performance of SDMs \citep{Fern2019}, motivating the development of approaches that go beyond purely environmental models. Improved understanding and prediction of species distributions are also important for conservation planning, including the identification and prioritization of areas for biodiversity protection and management.

In this context, JSDMs have been proposed as an alternative, aiming to model multiple species occurrence data simultaneously. Early contributions in this literature were developed mainly for presence–absence data \citep{Latimer2009,pollock2014}, while more recent work has extended these ideas to the presence-only case \citep{fithian2015, renner2019, botella2021, escamilla2022}. 
Overall, the literature on joint modeling for presence-only data is still relatively incipient, reflecting both the intrinsic limitations posed by sampling bias and the computational complexity of joint models \citep{escamilla2022}.

In this sense, the present work builds upon \citet{moreira2022analysis}, which proposes a univariate Bayesian model for presence-only data based on Poisson processes. To address the intractability of the likelihood, \citet{moreira2022analysis} introduce latent processes that produce an augmented likelihood that can be directly evaluated. An analogous strategy is adopted, allowing the likelihood to be evaluated exactly. Importantly, some of the latent processes carry a meaningful interpretation, representing unobserved occurrences. This augmentation is therefore not only instrumental for exact inference, but also plays a central role in the multivariate setting, where dependence between processes is modeled through both observed and unobserved points.

This work is also motivated by applications in the Amazon region, where presence-only data are common in both ecological and archaeological studies. In particular, the spatial distribution of certain tree species has been linked to archaeological indicators of pre-Columbian Indigenous land use (e.g., earthworks and dark earths), suggesting that the Amazon forest is not pristine, but shaped by long-term interactions with Indigenous peoples \citep{Thomas2015,levis2017,McMichael2025}. This paper presents a joint analysis of presence-only records for three socio-economically important tree species—\textit{Bertholletia excelsa}, \textit{Dipteryx odorata}, and \textit{Handroanthus serratifolius}—together with datasets of archaeological sites with earthworks and Amazonian Dark Earth locations. The central objective is to assess whether the current distributions of these species are associated with archaeological sites.

These applications highlight the need for statistical models that can estimate interpretable measures of relationships between presence-only datasets. A related analysis was conducted by \citet{peripato2023more}, where the distribution of earthworks in Amazonia was first estimated using a Poisson process model \citep{moreira2022analysis}, and the resulting predicted probabilities were then used as covariates in a second-stage generalized linear model to explain the occurrence of multiple tree species. While this plug-in approach provided initial evidence of association, it does not account for the uncertainty propagated from the first stage estimates to the second. The present work builds on this idea by proposing a joint model that fully accounts for uncertainty of all random quantities.

To represent potential relationships, the model is specified using Bayesian networks, where each point process corresponds to a node in a directed acyclic graph, and edges encode conditional dependence. The graph structure is assumed to be known and reflects prior knowledge about the problem under study, and inference is conducted conditional on this structure. Nevertheless, the proposed model also provides tools for graph estimation: inference on specific parameters allows the strength of each edge to be assessed and supports the removal of unsupported connections. Extensions to structure learning are also considered, enabling the graph to be estimated directly from the data.

Importantly, fully observed data can be viewed as a particular case of the proposed model for presence-only, corresponding to scenarios where sampling and observational biases are absent. In this case, the model reduces to a simpler formulation, as discussed later in the paper.

The rest of the paper is organized as follows. Section \ref{s:preli} introduces background material on multivariate Poisson processes and Bayesian networks. Section \ref{s:model} presents the proposed augmented joint model, along with its inferential aspects. Section~\ref{s:comput} provides an overview of the MCMC algorithm used for posterior inference. 
Section \ref{s:application} evaluates the methodology through simulation studies and presents an application to archaeological and ecological data from Amazonia. Section \ref{s:final} concludes with final remarks.

\section{Preliminaries}
\label{s:preli}

The observed point patterns are modeled in this work using Poisson processes (PPs), more specifically in a multivariate (or multitype) setting. Let $(X_1, \dots, X_N)$ denote a collection of point processes, where each component $X_i$ is a stationary Poisson point process defined on a common spatial domain $D \subset \mathbb{R}^p$, with $p \geq 1$, and intensity function $0 < \lambda_i < \infty$, for $i = 1, \ldots, N$.  
Throughout, we write \(X_i \sim PP(\lambda_i)\) to denote a Poisson process with intensity function \(\lambda_i\), $i=1,\dots,N$.
The index $i$ labels the different types of observed point patterns (e.g., distinct species) and does not imply any intrinsic ordering or hierarchy among them.

As in any Poisson point process models, the proposed joint model aims to estimate the intensity function associated with each process. When the PP components are mutually independent, constructing a joint model is straightforward, as each process can be modeled separately. 
However, in many practical situations, the processes are not independent, and the goal is to characterize the spatial association (positive or negative) between different types of events.

The multivariate point process formulation enables the analysis of two complementary aspects: (i) the relationships between distinct processes and (ii) the spatial intensity of each individual process.

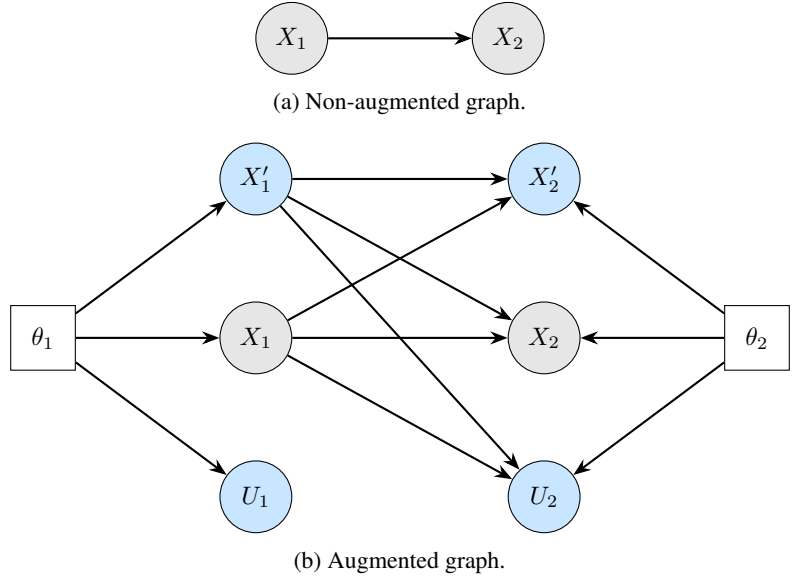
\begin{figure}[htbp]
\centering

\begin{subfigure}{0.9\columnwidth}
    \centering
    \begin{tikzpicture}[node distance=2cm, scale=0.95, transform shape]
        \node[observed] (X1) {\(X_1\)};
        \node[observed, right=of X1] (X2) {\(X_2\)};
        \draw[arrow] (X1) -- (X2);
    \end{tikzpicture}
    \caption{Non-augmented graph.}
\end{subfigure}

\vspace{0.3cm}

\begin{subfigure}{0.95\columnwidth}
    \centering
    \begin{tikzpicture}[
        node distance=1.2cm and 1.8cm,
        scale=0.95,
        transform shape
    ]

    \node[latent]                      (X1prime) {\(X_1'\)};
    \node[observed, below=of X1prime]  (X1)      {\(X_1\)};
    \node[latent,   below=of X1]       (UX1)     {\(U_{1}\)};
    \node[param,    left=of X1, xshift=-0.2cm] (Theta1) {\(\theta_1\)};

    \node[latent,  right=3cm of X1prime] (X2prime) {\(X_2'\)};
    \node[observed, below=of X2prime]    (X2)      {\(X_2\)};
    \node[latent,   below=of X2]         (UX2)     {\(U_{2}\)};
    \node[param,    right=of X2, xshift=0.2cm] (Theta2) {\(\theta_2\)};

    \draw[arrow] (X1) -- (X2);

    \draw[arrow] (Theta1) -- (X1);
    \draw[arrow] (Theta1) -- (X1prime);
    \draw[arrow] (Theta1) -- (UX1);

    \draw[arrow] (Theta2) -- (X2);
    \draw[arrow] (Theta2) -- (X2prime);
    \draw[arrow] (Theta2) -- (UX2);

    \draw[arrow] (X1prime) -- (X2);
    \draw[arrow] (X1prime) -- (X2prime);
    \draw[arrow] (X1prime) -- (UX2);
    \draw[arrow] (X1) -- (X2prime);
    \draw[arrow] (X1) -- (UX2);

    \end{tikzpicture}
    \caption{Augmented graph.}
\end{subfigure}

\caption{Bivariate example illustrating the augmented graphical structure. (a) A graph with a single directed edge \(X_1 \to X_2\). (b) The corresponding augmented graph, where gray nodes represent observed data, blue nodes correspond to latent processes, and rectangles indicate model parameters.}
\label{fig:bivariate-model-graph}
\end{figure}

To represent these dependence structures, Bayesian networks (BNs) are introduced, providing a way to encode conditional independence relationships between processes. Formally, a BN is defined as a pair $\mathcal{B} = (\mathcal{G}, P)$, where $\mathcal{G} = (\mathcal{X}, \mathcal{E})$ is a directed acyclic graph (DAG), with nodes $\mathcal{X} = \{X_1, \dots, X_N\}$ representing random quantities and directed edges $\mathcal{E}$ encoding dependencies, with $P$ denoting a probability distribution over $\mathcal{X}$ \citep{koller2009probabilistic}. The directionality of the edges implies that the model encodes asymmetric relationships. This is illustrated in Figure~\ref{fig:bivariate-model-graph}, where panel (a) represents a configuration in which $X_2$ depends on $X_1$, as indicated by the directed edge $X_1 \to X_2$.

For a given node $X_i$, $i=1,\dots,N$, the set of nodes with directed edges pointing to $X_i$ is called its parents, denoted by $\mathrm{Pa}(X_i)$. 
The (possibly empty) set of indices of the parents of $X_i$ is denoted by $J_i \subseteq \{1,\dots,N\}\setminus\{i\}$. 
A very useful property of BNs is that the joint distribution can be factorized as a product of conditional distributions, where each term corresponds to the distribution of a node given its parents:
\begin{equation}
P(x_1, \dots, x_N\mid \theta) = \prod_{i=1}^{N} P(x_i \mid \mathrm{Pa}(x_i),\theta),
\label{eq:bn_factorization}
\end{equation}
where $\theta$ is the collection of model parameters, to be specified in the next section.

In the proposed model, the graph $\mathcal{G}$ initially includes $N$ nodes, each representing an inhomogeneous Poisson process: $X_i \mid \mathrm{Pa}(x_i),\theta \sim PP(\lambda_i),\, i=1,\dots,N$, where the intensity $\lambda_i$ is specified as a function of $\theta$, the realization of the parent processes, and the covariates.

An important motivation for adopting BNs in this context lies in their interpretability. The DAG representation often reflects the intuitive understanding that domain experts have about their data, making it a natural tool for representing conditional independencies \citep{Krapu2023}. This enables the incorporation of expert knowledge about which processes are expected to influence others, in a manner that remains probabilistically manageable. For instance, if a process $X_j$ is expected to influence another process $X_i$, with $j \neq i$ and $i=1,\dots,N$, this is encoded by including $X_j \in \mathrm{Pa}(X_i)$.

\section{The joint model}
\label{s:model}

This section introduces the proposed joint model. Section~\ref{s:model_formulation} presents the model for multiple presence-only data, which constitutes the most general and challenging setting due to preferential sampling. Section~\ref{s:inference} describes the Bayesian inference procedures, and Section~\ref{s:structure_learning} introduces tools to estimate the dependence graph from the data. Section~\ref{s:presence_absence} then considers fully observed data as a special case of the presence-only formulation.

A baseline formulation for joint modeling the point processes introduced in Section~\ref{s:preli} is first described. For each process $X_i$, $i=1,\dots,N$, let $\lambda_i(s,\mathrm{Pa}(x_i))$ denote its intensity function at location $s \in D$, conditional on $\mathrm{Pa}(x_i)$ and $\theta$.

Under this formulation, the likelihood associated with the observed data $\mathscr{D} = \left(x_1,\dots,x_N\right)$ is given by
\begin{equation}
\mathcal{L}_{\mathscr{D}}(\theta) \propto \prod_{i=1}^{N}
\exp\left(-\int_{D} \lambda_i(s,\mathrm{Pa}(x_i))\, ds\right)
\prod_{s \in x_i} \lambda_i(s,\mathrm{Pa}(x_i)),
\label{eq:base_likelihood}
\end{equation}
where the factorization follows directly from Equation~(\ref{eq:bn_factorization}).

This likelihood poses important challenges. In particular, the integrals in Equation~(\ref{eq:base_likelihood}) are typically not available in closed form, a well-known difficulty in Poisson point process models. Here, this issue arises for $N$ integrals, all of which are intractable. A common approach is to approximate these integrals using numerical schemes, such as quadrature rules and spatial discretizations. While adequate in simpler settings, this becomes more challenging in multivariate models, where the accumulation of approximation errors can adversely affect both inference and predictive performance.

To address this issue, an augmented representation of the model is introduced, enabling exact Bayesian inference by avoiding the need for likelihood approximations. The key idea is to reformulate the model by introducing latent processes, leading to a tractable augmented likelihood expression.

\subsection{Augmented model formulation}
\label{s:model_formulation}

To overcome the intractability of the likelihood, a data augmentation strategy is adopted. This approach builds on \citet{goncalves2018} and \citet{moreira2022analysis}, which extended the idea to presence-only data.

For each $X_i$, $i=1,\dots,N$, two latent processes are introduced: $X_i'$, representing unobserved occurrences, and $U_i$, an auxiliary process required for the augmented construction, without direct interpretation.

Conditional on their parents, which are defined to be the same for the three nodes, the processes are specified as Poisson processes, $i=1,\dots,N$:
\begin{align}
X_i \mid \mathrm{Pa}(x_i), \theta &\sim PP\big(\lambda_i^* q_i(\cdot) p_i(\cdot)\big), \label{eq:Xi} \\
X_i' \mid \mathrm{Pa}(x_i), \theta &\sim PP\big(\lambda_i^* q_i(\cdot)(1-p_i(\cdot))\big), \label{eq:Xiprime} \\
U_i \mid \mathrm{Pa}(x_i), \theta &\sim PP\big(\lambda_i^*(1-q_i(\cdot))\big).\label{eq:Ui}
\end{align}

Here, $\lambda_i^*$, $i=1,\dots,N$, is a positive constant that acts as an upper bound for the intensity. The parametric forms of $q_i(\cdot)$ and $p_i(\cdot)$ are specified later.


The likelihood function for the model is augmented with the latent processes $L_i = (X_i', U_i)$, $i = 1, \dots, N$, which are grouped into $L = (L_1, \dots, L_N)$, yielding
\begin{equation}
\mathcal{L}_{\mathscr{D},L}(\theta) = \pi(\mathscr{D},L \mid \theta).
\end{equation}
Using the BN factorization in Equation~\eqref{eq:bn_factorization}, the augmented likelihood can be written as
\begin{equation}
\mathcal{L}_{\mathscr{D},L}(\theta)
= \prod_{i=1}^{N}
\pi(x_i \mid \mathrm{Pa}(x_i),\theta)\,
\pi(x_i' \mid \mathrm{Pa}(x_i),\theta)\,
\pi(u_i \mid \mathrm{Pa}(x_i),\theta).
\label{eq:aug_factorization}
\end{equation}

Substituting the Poisson process densities for each component gives
\begin{align}
\mathcal{L}_{\mathscr{D},L}(\theta)
&\propto
\exp\left(-|D|\sum_{i=1}^{N}\lambda_i^*\right)
\prod_{i=1}^{N}
\Bigg[
(\lambda_i^*)^{n_{x_i}+n_{x_i'}+n_{u_i}} \nonumber \\
&\times
\prod_{s\in x_i} q_i(s)p_i(s)
\prod_{s\in x_i'} q_i(s)(1-p_i(s))
\prod_{s\in u_i} (1-q_i(s))
\Bigg],
\label{eq:aug_lik_final}
\end{align}
where $|D|$ denotes the Lebesgue measure of the domain, equal to its area when $D \subset \mathbb{R}^2$.

Importantly, the intractable integrals in the original likelihood reduce to the term $|D|\sum_{i=1}^N \lambda_i^*$. This simplification is a direct consequence of the augmentation scheme, yielding a fully tractable likelihood. In particular, from the superposition of $X_i$, $X_i'$, and $U_i$, each conditional on $\mathrm{Pa}(x_i)$, whose intensities sum to $\lambda_i^*[q_i(s)p_i(s) + q_i(s)(1-p_i(s)) + (1-q_i(s))] = \lambda_i^*$, for all $i=1,\dots,N$, it follows that the superposed process is a homogeneous Poisson process with constant intensity $\lambda_i^*$.

Equivalently, this construction can be interpreted as a thinning mechanism: the superposed process $X_i \cup X_i'$, $i=1,\dots,N$, follows a Poisson process with intensity $\lambda_i^* q_i(\cdot)$, and $X_i$ corresponds to retained points with probability $p_i(\cdot)$.

The model specification is completed by defining $q_i(\cdot)$ and $p_i(\cdot)$, $i=1,\dots,N$. 
The function $q_i(\cdot)$, $i=1,\dots,N$, combines the effects of environmental covariates and dependence on other processes. In turn, $p_i(\cdot)$, $i=1,\dots,N$, can be interpreted as a detection probability, governing whether a potential occurrence is observed.  
These functions take values in $[0,1]$ and are defined through link functions $g_i$, for $i=1,\dots,N$:
\begin{align}
g_i\big[q_i(s)\big] &= Z_i(s)^\top \beta_i + d_i(s,\mathrm{Pa}(X_i))^\top \eta_i, \\
g_i\big[p_i(s)\big] &= W_i(s)^\top \delta_i.
\end{align}

Here, $Z_i(s)$ and $W_i(s)$, $i=1,\dots,N$, are covariate vectors for intensity (typically environmental variables) and observability, respectively, and $\beta_i$ and $\delta_i$ are the corresponding coefficient vectors.

The term $d_i(s,\mathrm{Pa}(X_i))$, $i=1,\dots,N$, is a vector-valued function summarizing spatial relationships between location $s$ and the parent processes. Possible choices include a minimum distance to parent points or counts within a given radius, allowing interpretable measures of interaction strength through the coefficient vector $\eta_i$, $i=1,\dots,N$.


In the Bayesian network considered, the graph $\mathcal{G}$ contains the set of nodes $\{X_i, X_i', U_i : i=1,\dots,N\}$. The augmented graph is defined such that (i) the triplet $(X_i, X_i', U_i)$, $\forall i=1,\dots,N$, contains no internal edges, (ii) all three nodes share the same parent set $\mathrm{Pa}(X_i) = \mathrm{Pa}(X_i') = \mathrm{Pa}(U_i)$, $i=1,\dots,N$, and (iii) parent sets include only observed processes and latent processes representing unobserved occurrences. More precisely, for each $i=1,\dots,N$, 
\[
\mathrm{Pa}(X_i) = \{\, X_j, X_j' : j \in J_i \,\}.
\]
Therefore, whenever a process $X_j$ influences $X_i$, $i=1,\dots,N$, both its observed and unobserved occurrences ($X_j$ and $X_j'$) are included in $\mathrm{Pa}(X_i)$.

Figure~\ref{fig:bivariate-model-graph} (b) illustrates the augmented structure in a bivariate setting. Each observed node is extended with its latent counterparts, and dependencies are inherited from the original graph. In practice, this implies that specifying the non-augmented graph is sufficient, as the corresponding relationships are automatically induced in the augmented representation.

In addition, processes~\eqref{eq:Xi}, \eqref{eq:Xiprime}, and \eqref{eq:Ui} are conditionally independent for all $i=1,\dots,N$, since in a BN any node is independent of its non-descendants given its parents \citep{koller2009probabilistic}. 

Importantly, the parent processes include the latent processes $X_i'$, $i=1,\dots,N$, which explicitly represent unobserved occurrences. This allows dependence between processes to be modeled by $d_i(\cdot)$, $i=1,\dots,N$, through their full (observed and unobserved) realizations. This is a desirable feature, as interactions are expected to arise from the complete underlying processes rather than from only the partial and potentially biased observed sample.

As a consequence, the inferential scheme must also include the estimation of the latent processes $X_i'$ and $U_i$, $i=1,\dots,N$. The model parameter set is given by
\begin{equation}
\theta = (\lambda^*, \beta, \delta, \eta),
\end{equation}
with $\lambda^* = (\lambda_1^*,\dots,\lambda_N^*)$, and similarly for \(\beta = (\beta_1,\dots,\beta_N)\), \(\delta = (\delta_1,\dots,\delta_N),\) and \(
\eta = (\eta_1,\dots,\eta_N)\). The full set of unknown quantities is therefore given by $\Theta = (\theta, L)$.

\subsection{Inference}
\label{s:inference}

Under the Bayesian approach, inference on unknown model quantities is obtained by combining prior information with the observed data. Inference is carried out for the full collection
\[
\Theta=(\theta,L)=(\lambda^*,\beta,\delta,\eta,L).
\]

In particular, the parameters \(\beta\) characterize how environmental covariates influence the underlying occurrence intensity, while \(\delta\) capture the effect of observability-related variables (e.g., proximity to roads or cities), allowing the model to account for sampling bias of presence-only data. The parameters \(\lambda^*\) control the overall scale of each process and directly affect the expected total number of points, including unobserved ones. Inference on the latent processes \(X_i'\), $i=1,\dots,N$, yields information about the number and spatial distribution of unobserved occurrences, which is often of primary interest in applications.

The inference is based on Bayes' theorem, which expresses the posterior distribution as
\begin{equation}
\pi(\Theta \mid \mathscr{D}) \propto \mathcal{L}_{\mathscr{D},L}(\theta)\,\pi(\theta),
\label{eq:bayes_inference}
\end{equation}
where \(\mathcal{L}_{\mathscr{D},L}(\theta)\) is the augmented likelihood and \(\pi(\theta)\) is the prior distribution for the model parameters. This formulation treats the latent processes \(L\) as unknown quantities to be estimated with \(\theta\), while also exploiting their role similar to the data $\mathscr{D}$ in the augmented likelihood.

Prior independence is assumed, $\pi(\theta)=\pi(\lambda^*)\pi(\beta,\eta)\pi(\delta)$, with independence also assumed across the parameters of each process. More precisely, for $i=1,\dots,N$, \(\lambda_i^*\) are assumed a priori independent, the vectors \(\delta_i\) are a priori independent across \(i\), and the concatenated vectors $\zeta_i = \begin{bmatrix} \beta_i \\ \eta_i \end{bmatrix}$ are also a priori independent across distinct processes. Gamma priors are assigned to each \(\lambda_i^*\), $i=1,\dots,N$, multivariate Normal priors to \(\zeta_i\), and multivariate Normal priors to \(\delta_i\). That is, \(\pi(\lambda_i^*) \sim \text{Gamma}(a_i,c_i)\), \(\pi(\zeta_i) \sim \mathcal{N}_k(b_i,B_i)\), and \(\pi(\delta_i) \sim \mathcal{N}_m(f_i,F_i)\), for \(i=1,\dots,N\). 
The joint prior for \((\beta_i,\eta_i)\) reflects the fact that both enter the same linear predictor for \(q_i(\cdot)\), for $i=1,\dots,N$.

The posterior distribution in Equation \eqref{eq:bayes_inference} is not available in closed form. Consequently, Markov chain Monte Carlo (MCMC) methods \citep{gamerman2006}, specifically a blocked Gibbs sampler, are used to obtain samples from the posterior. To improve computational efficiency, the unknown quantities are sampled in four blocks: $\zeta,\delta,L,$ and $\lambda^*$. Thus, each MCMC iteration consists of sampling from the full conditional distributions of \(\zeta\), \(\delta\), \(L\), and \(\lambda^*\).

\subsection{Learning the graph structure}
\label{s:structure_learning}

The inferential scheme presented is conditional on a fixed graph $\mathcal{G}$, which is assumed to encode the relevant dependence structure among the processes. In many applications, such a graph may be specified based on prior information and expert knowledge from the application domain. However, in the absence of prior information, it is natural to consider data-driven strategies for learning the graph structure.

A widely used approach to structure learning in graphical models relies on score-based methods, which compare candidate graphs through quantities such as the marginal likelihood \citep{koller2009probabilistic}. However, in the present setting, this strategy becomes computationally challenging, as each candidate graph requires fitting the full model via MCMC, which is particularly demanding for large $N$.

As an alternative, a simple two-stage strategy is proposed, taking advantage of the role of parameters $\eta$. Specifically, one may start from a sufficiently connected graph, including all edges deemed plausible from an application standpoint, and fit the model under this initial specification. The posterior inference on $\eta$ can then be used to assess the strength of each dependence and guide the removal of unsupported edges. In this sense, the parameters $\eta$ act as weights linking the graphical structure to the probabilistic model, quantifying the influence of parent processes.

In simple cases where each $\eta_i$ is scalar and corresponds to a single parent process, edge selection can be based on posterior credibility intervals. In particular, if the credibility interval for $\eta_i$ contains zero, this indicates a lack of evidence for the corresponding edge, suggesting that it may be removed. When $\eta_i$ is vector-valued, joint hypothesis tests for the nullity of all components may be employed to assess the presence of the corresponding dependency \citep{Migon2014}. 

An important consideration arises when $\eta_i$ encodes dependence on multiple parent processes simultaneously, for instance through summary measures computed over the union of parent point patterns. In such cases, a null estimate of $\eta_i$ does not allow one to remove the edge associated with any specific parent, as distinct effects may cancel out. For this reason, when the goal is structure learning, it is preferable to define terms $d(\cdot)$ that isolate the contribution of each parent process, even at the cost of increased computational time.

Finally, it is worth emphasizing that while unnecessary edges can be identified and removed according to the estimated parameters $\eta$, missing edges cannot be recovered if they are not included in the initial graph. Therefore, from a modeling perspective, it is generally preferable to begin with a more connected graph and subsequently prune unsupported dependencies.

\subsection{Modeling fully observed data}
\label{s:presence_absence}

A particular case of the proposed joint model is considered in which all events $X_i$, $i=1,\dots,N$, are fully observed. From a modeling perspective, this corresponds to setting $p_i(s)=1$ $\forall  s\in D$ and $i=1,\dots,N$, so that every event is detected with probability one. In this case, the observability component disappears and the model depends only on the functions $q_i(\cdot)$, $i=1,\dots,N$, as the sampling mechanism is no longer biased.

When the processes are fully observed, locations in $D$ that do not belong to the observed point patterns correspond to true absences. As a result, latent processes $X_i'$, $i=1,\dots,N$, which account for unobserved events in the presence-only setting, are no longer required. The model is then defined in terms of the observed processes $X_i$ and the latent processes $U_i$, for $i=1,\dots,N$, representing presences and absences, respectively:
\begin{align}
    X_i\mid \mathrm{Pa}(x_i),\theta &\sim PP\left(\lambda_i^*q_i(\cdot)\right),\\
    U_i\mid \mathrm{Pa}(x_i),\theta &\sim PP\left(\lambda_i^*(1-q_i(\cdot))\right).
\end{align}

As before, this construction can be interpreted as an independent thinning of a homogeneous Poisson process with intensity $\lambda_i^*$, with retention probability $q_i(s)$.

The likelihood function is then given by
\begin{equation}
\mathcal{L}_{\mathscr{D},U}(\theta)
\propto
\exp\left(-|D|\sum_{i=1}^{N}\lambda_i^*\right)
\prod_{i=1}^{N}
\Bigg[
(\lambda_i^*)^{n_{x_i}+n_{u_i}}
\prod_{s\in x_i} q_i(s)
\prod_{s\in u_i} (1-q_i(s))
\Bigg],
\label{eq:aug_observed}
\end{equation}
where $U=\{U_i: i=1,\dots,N\}$ denotes the collection of latent processes.

The augmented graph in this setting is obtained by simply removing the nodes $X_i'$, while preserving the same parent structure. In particular, for each $i=1,\dots,N$, the parent sets are given by $\mathrm{Pa}(X_i)=\mathrm{Pa}(U_i)=\{X_j: j \in J_i\}$.

\section{Computational aspects}
\label{s:comput}

This section derives the full conditional distributions, followed by an overview of the MCMC scheme used for posterior inference. The derivations are presented for the presence-only setting, as the fully observed case arises as a special case of this formulation.

Given prior independence, the full conditional distribution of \(\lambda^*\) factorizes as $\pi(\lambda^* \mid \cdot)=\prod_{i=1}^N \pi(\lambda_i^* \mid \cdot)$, and, by conjugacy with the augmented likelihood, the full conditional of \(\lambda_i^*\) is also Gamma distributed as $\pi(\lambda_i^*\mid\cdot) \sim \text{Gamma}\!\left(a_i+n_i,c_i+|D|\right)$, $i=1,\dots,N$, where \(n_i = n_{x_i} + n_{x_i'} + n_{u_i}\) is the total number of observed and latent points.

Likewise, the latent block factorizes as
\begin{equation}
    \pi(L\mid\cdot)=\prod_{i=1}^N \pi(L_i\mid\mathrm{Pa}(x_i),\theta),
\end{equation}
and for $i=1,\dots,N$, $\pi(L_i\mid\mathrm{Pa}(x_i),\theta)=\pi(X_i'\mid\mathrm{Pa}(x_i),\theta)\,\pi(U_i\mid\mathrm{Pa}(x_i),\theta)$ as follows from Equation~\eqref{eq:aug_factorization}.
From the augmented construction, the full conditional distributions of \(X_i'\) and \(U_i\), $i=1,\dots,N$, are Poisson processes with intensity functions given by Equations~\eqref{eq:Xiprime} and \eqref{eq:Ui}, respectively. Therefore, they can be sampled using a Poisson thinning algorithm \citep{lewis1979}. 

For the regression coefficients \(\zeta\) and \(\delta\), the assumption of prior independence implies that their full conditional distributions factorize over \(i = 1,\ldots,N\). Hence, $\pi(\zeta\mid\cdot) = \prod_{i=1}^{N}\pi(\zeta_i\mid\cdot)$ and $\pi(\delta\mid\cdot) = \prod_{i=1}^{N}\pi(\delta_i\mid\cdot)$. In both cases, the inferential scheme relies on a binary representation in which points can be seen as successes or failures. For sampling the coefficients \(\zeta_i\), the points in \(x_i \cup x_i'\) are treated as successes, whereas the points in \(u_i\) are treated as failures. For \(\delta_i\), the points in \(x_i\) and \(x_i'\) play the roles of successes and failures, respectively.

The resulting conditional distributions depend on the chosen link function. Under the logit link, inference is performed through the Pólya--Gamma augmentation scheme of \citet{polson2012}, while under the probit link the Gaussian construction of \citet{albert1993} is adopted. 
The complete derivations and sampling details are provided in Appendix~\ref{appendix:coefs}.


Algorithm \ref{algo:MCMC_full} summarizes the complete MCMC procedure for the proposed model. A key aspect of the implementation is the use of a topological ordering of the graph $\mathcal{G}$, that is, an ordering such that all parent processes of a given node are sampled before the node itself. The use of a topological ordering ensures the validity of the Gibbs sampler, as each conditional distribution depends on the values of the parent processes. In particular, if $\mathcal{G}$ contains any directed cycles, no such ordering exists, which is consistent with the requirement that $\mathcal{G}$ must be a DAG. The MCMC algorithm is then run for a sufficiently large number of iterations until convergence is achieved.

\begin{algorithm}[htbp]
\caption{MCMC scheme for the proposed joint model with presence-only data} 
\label{algo:MCMC_full}

Initialize ${\lambda^*}^{(0)}$, $\beta^{(0)}$, $\delta^{(0)}$, $\eta^{(0)}$\;
Compute a topological ordering of $\mathcal{G}$, denoted by $\tau=(\tau_1,\ldots,\tau_N)$\;

\For{$t = 1$ \KwTo number of iterations}{

    \For{$k = 1$ \KwTo $N$}{
        $i \leftarrow \tau_k$\;
        
        \tcp{Sample latent processes $L$}
        Sample $S \sim \text{Poisson}(\lambda_i^* |D|)$ and distribute points uniformly over $D$\;
        
        \ForEach{$s \in S$}{
            Compute $d_i(s,\mathrm{Pa}(X_i))$, $q_i(s)$ and sample $V \sim \text{Uniform}(0,1)$\;
            \eIf{$V > q_i(s)$}{
                Assign $s$ to $U_i$\;
            }{
                Compute $p_i(s)$\;
                \eIf{$V > q_i(s)p_i(s)$}{
                    Assign $s$ to $X_i'$\;
                }{
                    Discard $s$\;
                }
            }
        }
    }

    \For{$k = 1$ \KwTo $N$}{
        $i \leftarrow \tau_k$\;

        \tcp{Sample $\lambda_i^*$}
        Sample $\lambda_i^* \sim \text{Gamma}\!\left(a_i + n_i, c_i + |D|\right)$\;

        \tcp{Sample regression coefficients*}

        \uIf{logit link}{
            
            \tcp{Intensity coefficients $\zeta_i$}
            Sample $\omega_{j}\mid\cdot \sim \text{Pólya\text{-}Gamma}\!\left(1,\tilde{Z}_i(s)^{\top}\zeta_i\right)$ for $j=1,\ldots,n_i$\;
            Sample $\zeta_i\mid\cdot \sim \mathcal{N}_k\!\left(m_{\zeta_i},V_{\zeta_i}\right)$\;
            
            \tcp{Observability coefficients $\delta_i$}
            Sample $\omega_{j}\mid\cdot \sim \text{Pólya\text{-}Gamma}\!\left(1,W_i(s)^{\top}\delta_i\right)$ for $j=1,\ldots,n_{\tilde{x}_i}$\;
            Sample $\delta_i\mid\cdot \sim \mathcal{N}_m\!\left(m_{\delta_i},V_{\delta_i}\right)$\;
        }
        \ElseIf{probit link}{
            
            \tcp{Intensity coefficients $\zeta_i$}
            Sample $\psi_{j}\mid\cdot \sim \mathcal{N}\!\left(\tilde{Z}_i(s)^{\top}\zeta_i,1\right)$, truncated at $0$ according to $y_{ij}$\;
            Sample $\zeta_i\mid\cdot \sim \mathcal{N}_k\!\left(\tilde{b}_i,\tilde{B}_i\right)$\;
            
            \tcp{Observability coefficients $\delta_i$}
            Sample $\psi_{j}\mid\cdot \sim \mathcal{N}\!\left(W_i(s)^{\top}\delta_i,1\right)$, truncated at $0$ according to $y_{ij}$\;
            Sample $\delta_i\mid\cdot \sim \mathcal{N}_m\!\left(\tilde{f}_i,\tilde{F}_i\right)$\;
        }
    }

    Store $L^{(t)}$, ${\lambda^*}^{(t)}$, $\beta^{(t)}$, $\delta^{(t)}$, $\eta^{(t)}$\;
    
}
\vspace{0.3cm} 
\footnotesize *Details are provided in Appendix~\ref{appendix:coefs}.
\end{algorithm}

\section{Model application}
\label{s:application}

This section presents both simulation studies and a real-data application in order to evaluate the proposed model. Section~\ref{sec:simul} investigates the inferential performance of the methodology under controlled simulation settings, while Section~\ref{sec:app} illustrates its applicability through an analysis of archaeological sites and species occurrence data in Amazonia.

\subsection{Simulation studies}
\label{sec:simul}

Two simulation studies were conducted to assess whether the proposed inferential scheme is capable of recovering the true model parameters under different dependence structures. All simulated datasets were generated within the unit square, which served as the spatial domain $D$. For each simulation study, four parameter scenarios were considered, and for every scenario 50 independent datasets were generated. A logit link was adopted for both the intensity and observability components.

Each point process was simulated using one intensity covariate and one observability covariate. Covariates were generated from independent Gaussian Processes over the unit square to mimic spatial patterns commonly observed in environmental variables. For all scenarios, $d$ was defined using a single distance covariate,
\begin{equation}
d(s, \text{Pa}(X_i)) = \min_{u \in \text{Pa}(X_i)} ||s-u||,
\end{equation}
corresponding to the minimum Euclidean distance between location $s$ and the set of locations from the parent processes.

Inference for each simulated dataset was based on a single MCMC chain with 70,000 iterations. The first 20,000 iterations were discarded as burn-in, and thinning was applied to the remaining samples by retaining one draw every 50 iterations, resulting in 1,000 posterior draws for each dataset. Trace plots of the log-posterior densities for all simulation studies and scenarios are provided in Figures S2--S9 in Supplementary Material, indicating convergence.

Prior distributions were specified as follows. The coefficients $\beta$, $\eta$, and $\delta$ received independent Normal priors with mean 0 and variance 10. The parameters $\lambda^*$ were assigned independent Gamma priors with shape and rate equal to 0.001. These weakly informative priors were kept fixed across all scenarios and simulations.


The first simulation study corresponds to the model illustrated in Figure~\ref{fig:bivariate-model-graph}, in which an observed process $X_1$ directly influences another process $X_2$, i.e., $X_1 \rightarrow X_2$. Four parameter scenarios were considered, with parameter values provided in Table~S1 of the Supplementary Material. The scenarios were designed to produce increasing proportions of unobserved occurrences in the latent process $X_1'$, thereby reducing the number of observed points in $X_1$. Across the 50 replications, the mean numbers of observed occurrences in $X_1$ were 554, 445, 269, and 129 for Scenarios 1 to 4, respectively. The corresponding mean proportions of unobserved occurrences, measured as the proportion of $X_1'$ within $X_1 \cup X_1'$, were 23.3\%, 38.2\%, 62.6\%, and 82.1\%.

This study was designed to evaluate the impact of increasing levels of unobserved data on parameter estimation, particularly for the parameter $\eta$, whose estimation depends on both the observed locations and the inferred $X'$ occurrences from the parent process.

Results are summarized in Table~\ref{tab:sim-coverage} through empirical coverage rates, defined as the proportion of times the true parameter value was contained within the corresponding 90\% credibility interval across the 50 replications. The corresponding credibility intervals are also displayed in Figures S10--S13 of the Supplementary Material.

\begin{table}[htbp]
\centering
\caption{Empirical coverage rates of the 90\% credibility intervals for all model parameters in simulation study 1 across the four scenarios. Each entry reports the percentage of the 50 simulated datasets in which the true parameter value was contained within the corresponding interval.}
\vspace{0.1cm}
\small
\begin{tabular}{c|c|cccc}
\hline
\multirow{2}{*}{Process} & \multirow{2}{*}{Parameter} & \multicolumn{4}{c}{Scenario} \\
                         &                            & 1 & 2 & 3 & 4 \\
\hline

\multirow{5}{*}{$X_1$}
    & $\beta_0$      & 98.0\%  & 86.0\%  & 94.0\%  & 94.0\% \\
    & $\beta_1$      & 100.0\% & 90.0\%  & 98.0\%  & 96.0\% \\
    & $\delta_0$     & 98.0\%  & 92.0\%  & 98.0\%  & 90.0\% \\
    & $\delta_1$     & 100.0\% & 94.0\%  & 96.0\%  & 90.0\% \\
    & $\lambda^*$    & 100.0\% & 96.0\%  & 96.0\%  & 94.0\% \\[0.3em]

\hdashline[0.4pt/5pt]

\multirow{6}{*}{$X_2$}
    & \rule{0pt}{1.4em}$\beta_0$ & 78.0\%  & 90.0\%  & 66.0\%  & 58.0\% \\
    & $\beta_1$                  & 96.0\%  & 98.0\%  & 90.0\%  & 90.0\% \\
    & $\eta$                     & 50.0\%  & 90.0\%  & 98.0\%  & 98.0\% \\
    & $\delta_0$                 & 100.0\% & 100.0\% & 100.0\% & 100.0\% \\
    & $\delta_1$                 & 98.0\%  & 98.0\%  & 98.0\%  & 100.0\% \\
    & $\lambda^*$                & 98.0\%  & 96.0\%  & 88.0\%  & 88.0\% \\

\hline
\end{tabular}
\label{tab:sim-coverage}
\end{table}

Overall, most parameters were accurately estimated across all scenarios. Empirical coverage rates were generally close to or slightly above the nominal 90\% level, indicating that the proposed inferential procedure provides reliable uncertainty quantification, although credibility intervals were, in some cases, slightly wider than necessary.

Particular attention was given to the parameter \(\eta\). The best performance was observed in Scenario 2, which achieved empirical coverage exactly equal to the nominal level (90\%) with an average of 445 observed occurrences in \(X_1\). As the proportion of unobserved occurrences increased (Scenarios 3 and 4), coverage rose to 98\%, suggesting increased estimation uncertainty and consequently wider credibility intervals. 
Scenario 1 yielded a lower empirical coverage rate (50\%) despite having the largest number of observed occurrences in \(X_1\). However, Figure S10 in the Supplementary Material shows that the corresponding credibility intervals remained entirely within the negative range, correctly recovering the direction of $\eta=-1$. Thus, the model still provided consistent inference regarding the sign of \(\eta\).

From a practical perspective, the results indicate that the parameter $\eta$ can be estimated reliably even when a substantial proportion of the parent process is unobserved. In particular, Scenario 4, with only approximately 129 observed occurrences in \(X_1\), still produced posterior means close to the true value \(\eta=-1\) (Figure S13 Supplementary Material).

One limitation observed in the simulation study is that a smaller number of observed occurrences in the parent process appears to affect the estimation of the intercept parameter \(\beta_0\). For process \(X_2\), coverage for this parameter decreased to 66\% and 58\% in Scenarios 3 and 4, respectively.


The second simulation study considers a more complex setting involving three observed point processes, with dependence relationships given by $X_1 \rightarrow X_2 \rightarrow X_3$. The corresponding augmented graph is provided in Figure~S1 of the Supplementary Material and the parameter values used to generate the simulated datasets are reported in Table~S2. In this study, selected edges were removed across the four simulated scenarios by setting the corresponding parameter to $\eta = 0$. The main objective was therefore to evaluate whether the proposed model is capable of correctly identifying the absence of relationships between processes through the estimation of $\eta$.

As in the first simulation study, 50 independent datasets were generated for each scenario, using the same prior specifications and MCMC settings previously described. Results are summarized in Table~S3 of the Supplementary Material through the empirical coverage rates of the 90\% credibility intervals. The corresponding intervals are shown in Figures S14--S17.

Particular attention is given to scenarios in which the true parameter was set to \(\eta = 0\). For process \(X_2\), the edge \(X_1 \rightarrow X_2\) was removed in Scenarios 2 and 4, resulting in empirical coverage rates of 94\% and 100\% for $\eta$, respectively. Similarly, for process \(X_3\), the edge \(X_2 \rightarrow X_3\) was removed in Scenarios 1 and 4, yielding coverage rates of 92\% and 94\%.

These results indicate that the proposed inferential scheme is capable of correctly identifying independence between processes. In scenarios where no interaction was present, the corresponding credibility intervals for \(\eta\) contained the null value with frequencies close to or above the nominal level. Consequently, this result suggests that the model is unlikely to infer $\eta \neq 0$ when the processes are in fact independent.

\subsection{Modeling archaeological and ecological processes in Amazonia}
\label{sec:app}

Archaeological evidence, such as earthworks and Amazonian Dark Earth (ADE) sites, indicates widespread and persistent Indigenous modification of the landscape in the Amazon forest. This raises important questions about the extent to which these past Indigenous management have influenced the forest composition. In particular, the spatial distribution of modern tree species may reflect past management practices, rather than purely environmental factors. Therefore, this application is motivated by the need to better understand the relationship between archaeological indicators of Indigenous land use and the current distribution of relevant tree species.

The model is applied to presence-only data on earthworks sites, ADE sites, and three tree species: \textit{H. serratifolius}, \textit{B. excelsa}, and \textit{D. odorata}. Earthworks are archaeological sites associated with anthropogenic landscape modifications, such as roads, fortified settlements, and geoglyphs \citep{WinklerPrins2021, peripato2023more}. Amazonian Dark Earths (ADEs), also known as \textit{terra preta}, are anthropogenic soil formations characterized by high nutrient content relative to surrounding Amazonian soils, and are widely interpreted as evidence of long-term pre-Columbian settlements \citep{mcmichael2014, WinklerPrins2021}. Species occurrences are specified as depending on the archaeological processes, with the aim of assessing whether pre-Columbian Indigenous land use may have contributed to shaping the present occurrence of these species.

These species were selected due to their socio-economic, ecological, and conservation relevance in Amazonia, as well as their distinct forms of human use. \textit{B. excelsa} is widely harvested for its edible nuts and represents an important source of income for local communities, whereas \textit{D. odorata} is valued for its aromatic seeds used in perfumes and flavorings. In contrast, \textit{H. serratifolius} is primarily exploited for its dense and highly valuable hardwood. Additionally, both \textit{H. serratifolius} and \textit{B. excelsa} are currently classified by the IUCN Red List as endangered \citep{IUCN_Hserratifolius_2021} and vulnerable species \citep{IUCN_Bexcelsa_1998}, respectively, further highlighting their conservation importance.

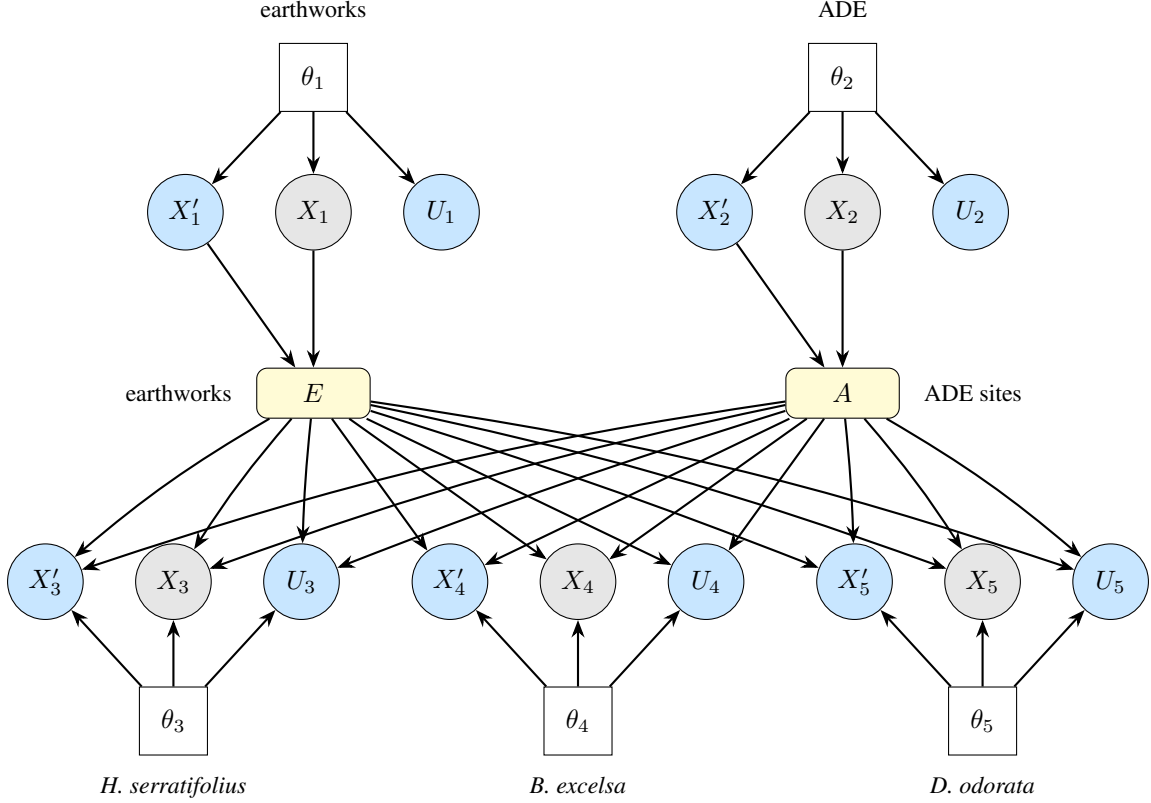
\begin{figure}[htbp]
\centering
\begin{tikzpicture}[
    node distance=0.58cm and 0.68cm,
    deterministic/.style={draw, rounded corners, fill=yellow!20, inner sep=6pt, minimum width=1.5cm}
]


\node[latent]                          (X3prime) {\(X_3'\)};
\node[observed, right=of X3prime]      (X3)      {\(X_3\)};
\node[latent, right=of X3]             (UX3)     {\(U_3\)};
\node[param, below=0.88cm of X3]       (Theta3)  {\(\theta_3\)};
\node[below=1.95cm of X3, font=\small] {\textit{H. serratifolius}};

\node[latent, right=0.95cm of UX3]     (X4prime) {\(X_4'\)};
\node[observed, right=of X4prime]      (X4)      {\(X_4\)};
\node[latent, right=of X4]             (UX4)     {\(U_4\)};
\node[param, below=0.88cm of X4]       (Theta4)  {\(\theta_4\)};
\node[below=1.95cm of X4, font=\small] {\textit{B. excelsa}};

\node[latent, right=0.95cm of UX4]     (X5prime) {\(X_5'\)};
\node[observed, right=of X5prime]      (X5)      {\(X_5\)};
\node[latent, right=of X5]             (UX5)     {\(U_5\)};
\node[param, below=0.88cm of X5]       (Theta5)  {\(\theta_5\)};
\node[below=1.95cm of X5, font=\small] {\textit{D. odorata}};


\node[deterministic, above=1.65cm of X4, xshift=-3.5cm] (E) {\(E\)};
\node[left=0.2cm of E, font=\small, align=right] {earthworks};

\node[deterministic, above=1.65cm of X4, xshift=3.5cm] (A) {\(A\)};
\node[right=0.2cm of A, font=\small, align=left] {ADE sites};


\node[observed, above=1.55cm of E] (X1) {\(X_1\)};
\node[latent, left=of X1]                          (X1prime) {\(X_1'\)};
\node[latent, right=of X1]                         (UX1) {\(U_1\)};
\node[param, above=0.82cm of X1]                   (Theta1) {\(\theta_1\)};
\node[above=1.95cm of X1, font=\small]             {earthworks};

\node[observed, above=1.55cm of A] (X2) {\(X_2\)};
\node[latent, left=of X2]                           (X2prime) {\(X_2'\)};
\node[latent, right=of X2]                          (UX2) {\(U_2\)};
\node[param, above=0.82cm of X2]                    (Theta2) {\(\theta_2\)};
\node[above=1.95cm of X2, font=\small]              {ADE};

\draw[arrow] (Theta1) -- (X1prime);
\draw[arrow] (Theta1) -- (X1);
\draw[arrow] (Theta1) -- (UX1);

\draw[arrow] (Theta2) -- (X2prime);
\draw[arrow] (Theta2) -- (X2);
\draw[arrow] (Theta2) -- (UX2);

\draw[arrow] (Theta3) -- (X3prime);
\draw[arrow] (Theta3) -- (X3);
\draw[arrow] (Theta3) -- (UX3);

\draw[arrow] (Theta4) -- (X4prime);
\draw[arrow] (Theta4) -- (X4);
\draw[arrow] (Theta4) -- (UX4);

\draw[arrow] (Theta5) -- (X5prime);
\draw[arrow] (Theta5) -- (X5);
\draw[arrow] (Theta5) -- (UX5);


\draw[arrow] (X1prime) -- (E);
\draw[arrow] (X1)      -- (E);

\draw[arrow] (X2prime) -- (A);
\draw[arrow] (X2)      -- (A);


\draw[arrow] (A) to[bend right=4] (X3);
\draw[arrow] (A) to[bend right=5] (X3prime);
\draw[arrow] (A) to[bend right=2] (UX3);

\draw[arrow] (E) to[bend right=4] (X3);
\draw[arrow] (E) to[bend right=5] (X3prime);
\draw[arrow] (E) to[bend right=2] (UX3);

\draw[arrow] (A) -- (X4);
\draw[arrow] (A) -- (X4prime);
\draw[arrow] (A) -- (UX4);

\draw[arrow] (E) -- (X4);
\draw[arrow] (E) -- (X4prime);
\draw[arrow] (E) -- (UX4);

\draw[arrow] (A) to[bend left=4] (X5);
\draw[arrow] (A) to[bend left=2] (X5prime);
\draw[arrow] (A) to[bend left=5] (UX5);

\draw[arrow] (E) to[bend left=4] (X5);
\draw[arrow] (E) to[bend left=2] (X5prime);
\draw[arrow] (E) to[bend left=5] (UX5);

\end{tikzpicture}
\caption{Application augmented graph. The deterministic nodes $E$ and $A$ represent the union of observed and unobserved earthworks and ADE sites, respectively. Nodes in light gray represent observed data, nodes in light blue correspond to latent processes, and the yellow node denotes the union of processes. White rectangles represent model parameters.}
\label{fig:application-graph-aug}
\end{figure}

Figure~\ref{fig:application-graph-aug} presents the augmented graph considered in the application. In this graph, the earthworks and ADE processes are represented by $X_1$ and $X_2$, respectively, while species occurrences are represented by $X_3$, $X_4$, and $X_5$, corresponding to \textit{H. serratifolius}, \textit{B. excelsa}, and \textit{D. odorata}. Deterministic nodes are introduced for earthworks and ADEs, with \(E = X_1 \cup X_1'\) and \(A = X_2 \cup X_2'\). This enables species associations to be evaluated with respect to the complete realization (observed and unobserved locations) of each archaeological process.

The graph structure encodes the assumption that the occurrence patterns of the selected species may be associated with the archaeological sites. In this formulation, the proposed model is used to assess whether such relationships are supported by the data, as well as to estimate their strength through the parameters \(\eta\).

Occurrences of earthworks and ADE sites are the same as those used in \citet{Walker2023}, with spatial data available in \citet{Walker_2023_Zenodo}. Species records were obtained from the Global Biodiversity Information Facility (GBIF) \citep{GBIFBackbone2023} and speciesLink \citep{speciesLink2025}, and subjected to standard cleaning procedures, including taxonomic standardization, filtering by coordinate precision, removal of records near centroids and biological institutions, and exclusion of duplicates.

\begin{figure}[htbp]
    \centering

    \begin{subfigure}{0.38\textwidth}
        \centering
        \includegraphics[width=\textwidth]{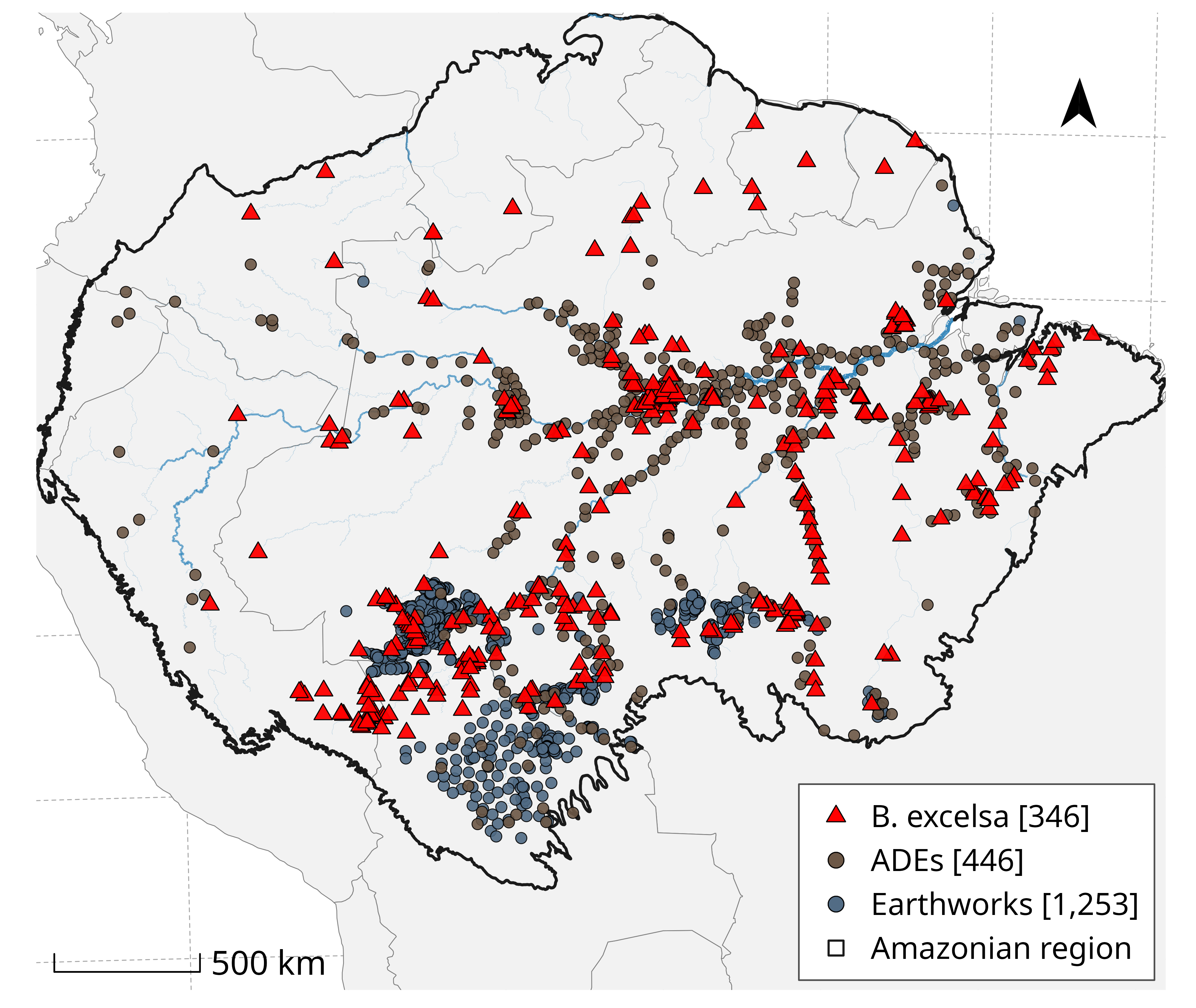}
        \caption*{\textbf{(A)}}
    \end{subfigure}
    \hspace{0.01\textwidth}
    \begin{subfigure}{0.38\textwidth}
        \centering
        \includegraphics[width=\textwidth]{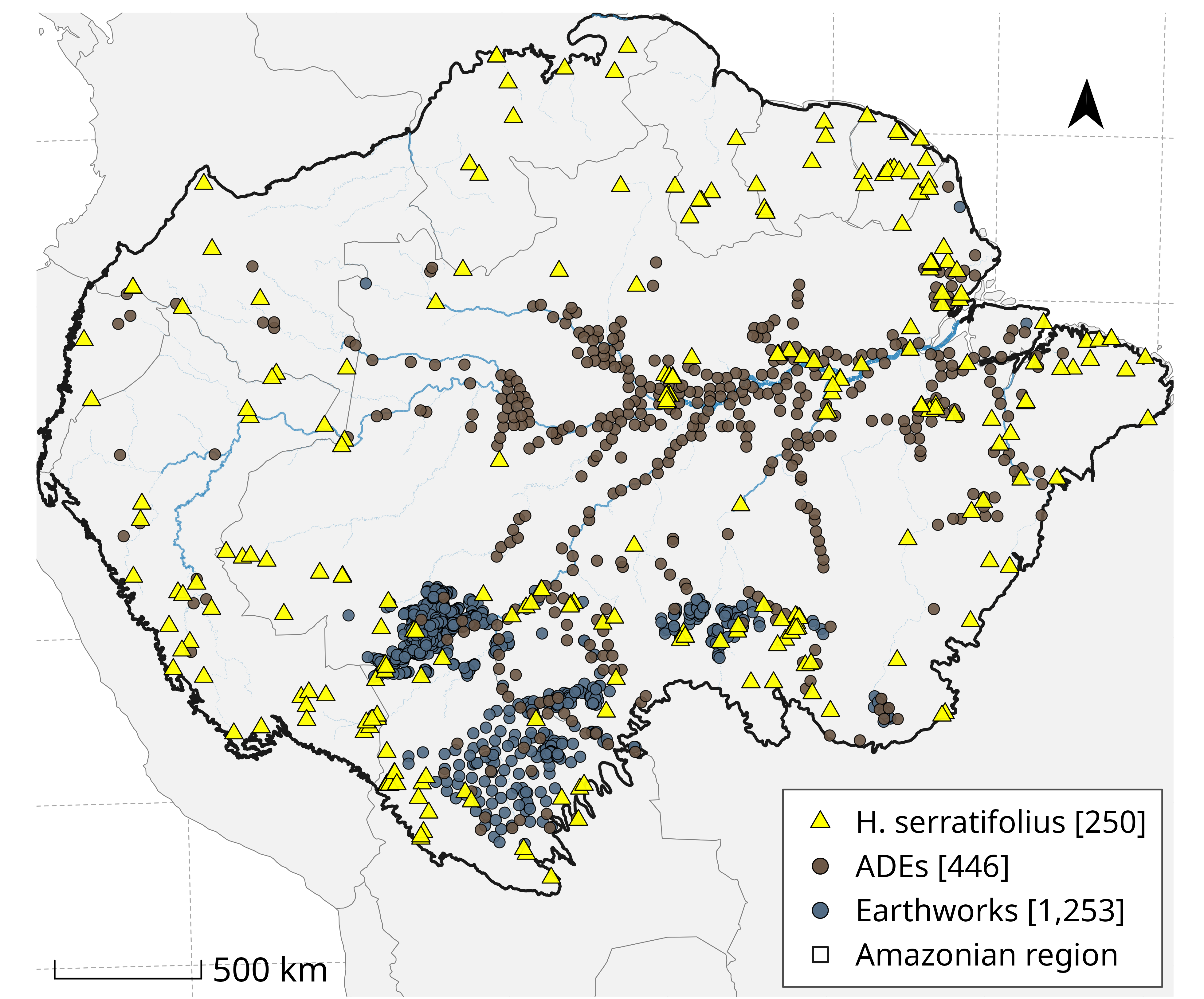}
        \caption*{\textbf{(B)}}
    \end{subfigure}

    \vspace{0.1cm}

    \begin{subfigure}{0.38\textwidth}
        \centering
        \includegraphics[width=\textwidth]{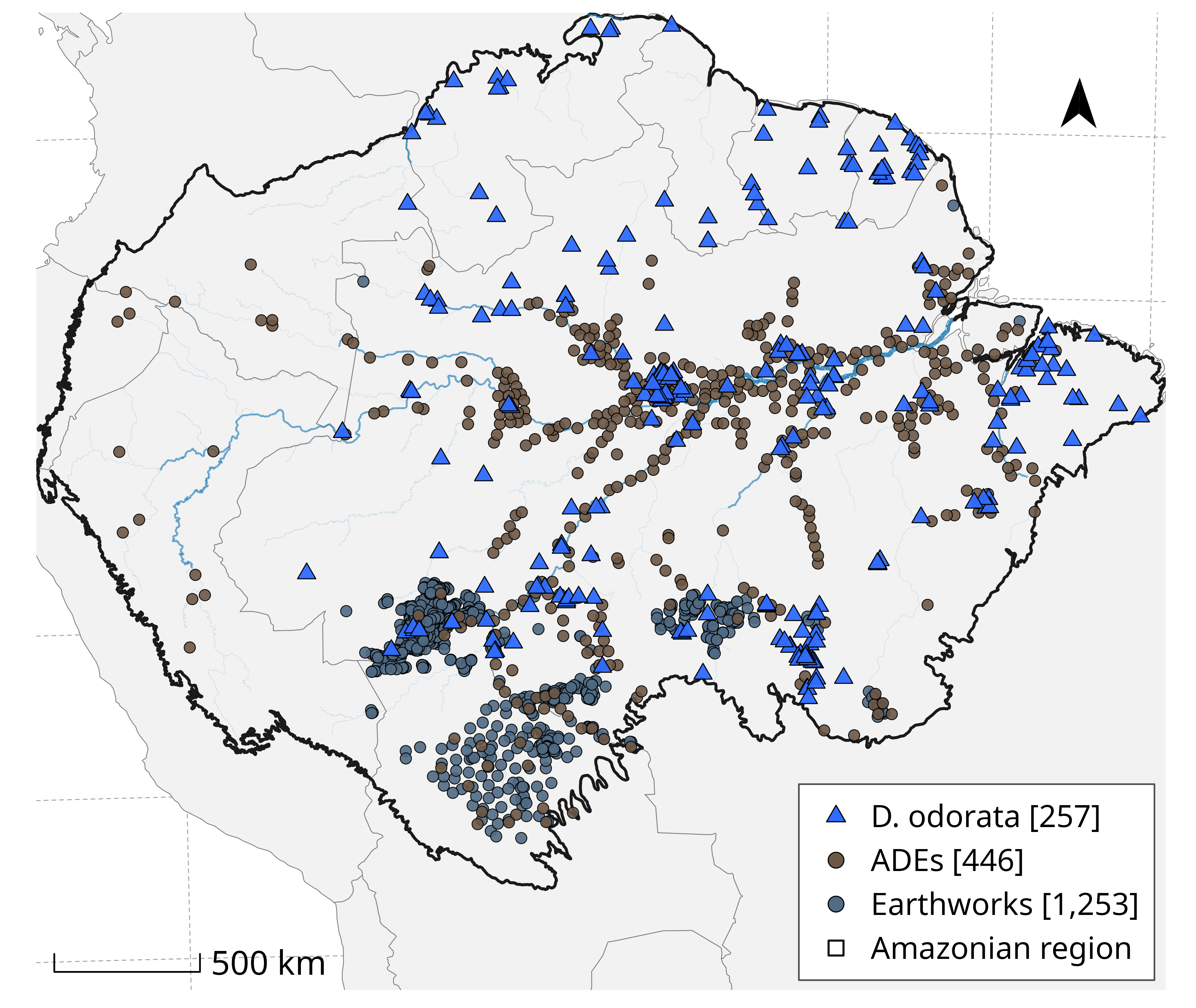}
        \caption*{\textbf{(C)}}
    \end{subfigure}

    \caption{Presence-only occurrences across the Amazonian region. Panels (A), (B), and (C) display the locations of \textit{B. excelsa}, \textit{H. serratifolius}, and \textit{D. odorata}, respectively. Triangles indicate species occurrence records, while circular markers indicate ADEs and earthworks locations. Major rivers are shown as blue lines.}
    \label{fig:panel_maps}
\end{figure}

All datasets were restricted to the extent of Amazonia, defining the study region \(D\) (approximately 6.7 million km\(^2\)). The final dataset comprises 1,253 earthwork sites, 446 ADE sites, 346 \textit{B. excelsa} locations, 257 \textit{D. odorata} locations, and 250 \textit{H. serratifolius} locations distributed across the region. Figure~\ref{fig:panel_maps} displays the spatial distribution of all occurrence records considered in the analysis.

Two observability covariates were considered for all datasets: tree cover and distance to roads. For each process, two intensity covariates were selected based on previous studies on similar data, prioritizing those identified as most relevant. For the species data, covariate choices were informed by \citet{Tourne2019} for \textit{B. excelsa}, \citet{Vitorino2016} for \textit{H. serratifolius}, and \citet{Carvalho2025} for \textit{D. odorata}. The covariates comprise bioclimatic, edaphic, hydrographic, and topographic variables. A complete description of the covariates and their data sources is provided in Table~S4 in the Supplementary Material. 
All covariates were standardized for model fitting.


A central focus of this application lies in the estimation of the parameters \(\eta\), which quantify the relationship between tree species occurrences and archaeological sites. The term \(d_i(\cdot)\), $i=3,4,5$, is specified as a vector-valued function with two components, capturing the proximity to earthworks and ADE sites separately.

Let \(r_{i1}(s)\) and \(r_{i2}(s)\) denote the minimum Euclidean distances from location \(s\) to the sets of earthworks or ADE locations, respectively:
\begin{equation}
    r_{ij}(s) = \min_{u \in X_j \cup X_j'} \| s - u \|, \quad j=1,2,
    \quad i=3,4,5.
\end{equation}

The term \(d_i(s) = \left( d_{i1}(s),\, d_{i2}(s) \right)\) is given by
\begin{equation}
    d_{ij}(s) = \frac{1}{r_{ij}(s)+0.1} \,\mathds{1}\{r_{ij}(s) < 25\}, \quad j=1,2,
    \quad i=3,4,5,
    \label{eq:distance_app}
\end{equation}

so that each component captures the truncated inverse distance to the nearest earthwork or ADE site, respectively. Distances greater than 25~km are assigned a zero effect. The constant \(0.1\) serves to prevent division by zero.

This threshold is motivated by \citet{levis2017}, which indicates that the influence of past human activity on forest composition decreases markedly beyond 25~km from archaeological sites. In addition, truncation reduces the computational cost of the MCMC algorithm by limiting the number of distance evaluations required at each iteration.

The parameters \(\eta_i = (\eta_{i1}, \eta_{i2})\) act on the components of \(d_i(s)\), allowing the effects of proximity to earthworks and ADE sites to be estimated separately. Positive values indicate an increase in the intensity of the species process as proximity to archaeological sites increases, whereas negative values indicate lower occurrence near such sites.


Normal priors with mean zero and variance 10 were assigned to the coefficients \(\beta\), \(\eta\), and \(\delta\). The intensity bounds \(\lambda^*\) were assigned Gamma priors with shape and rate equal to 0.001 (variance of 1{,}000), corresponding to weak prior information.


Two MCMC chains were run for 130{,}000 iterations each. The first 105{,}000 iterations were discarded as burn-in, and the remaining samples were thinned by retaining every 25th draw, yielding 1{,}000 posterior samples per chain (2{,}000 in total). The trace plots indicated convergence and are provided in Figure~S18 in the Supplementary Material.

\begin{figure}[htbp]
\centering

\begin{subfigure}[c]{0.48\textwidth}
\centering
\resizebox{\textwidth}{!}{%
\begin{tikzpicture}[node distance=2cm and 3.5cm, ->, thick, every node/.style={align=center}]

\def\r{0.75cm}
\def\d{1.6cm}
\def\densw{1.02cm}
\definecolor{myblue}{RGB}{25, 113, 158}

\tikzset{
  posedge/.style={->, thick, draw=myblue},
  negedge/.style={->, thick, draw=red!75!black},
  nulledge/.style={->, thick, draw=gray!90, dashed}
}

\node (A) at (0,1.5) {%
  \begin{tikzpicture}
    \clip (0,0) circle (\r);
    \node at (0,0) {\includegraphics[width=\d,height=\d]{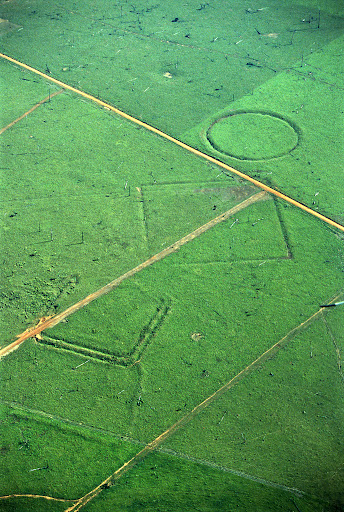}};
    \draw[black, thick] (0,0) circle (\r);
  \end{tikzpicture}
};

\node (B) at (0,-1.5) {%
  \begin{tikzpicture}
    \clip (0,0) circle (\r);
    \node at (0,0) {\includegraphics[width=\d,height=\d]{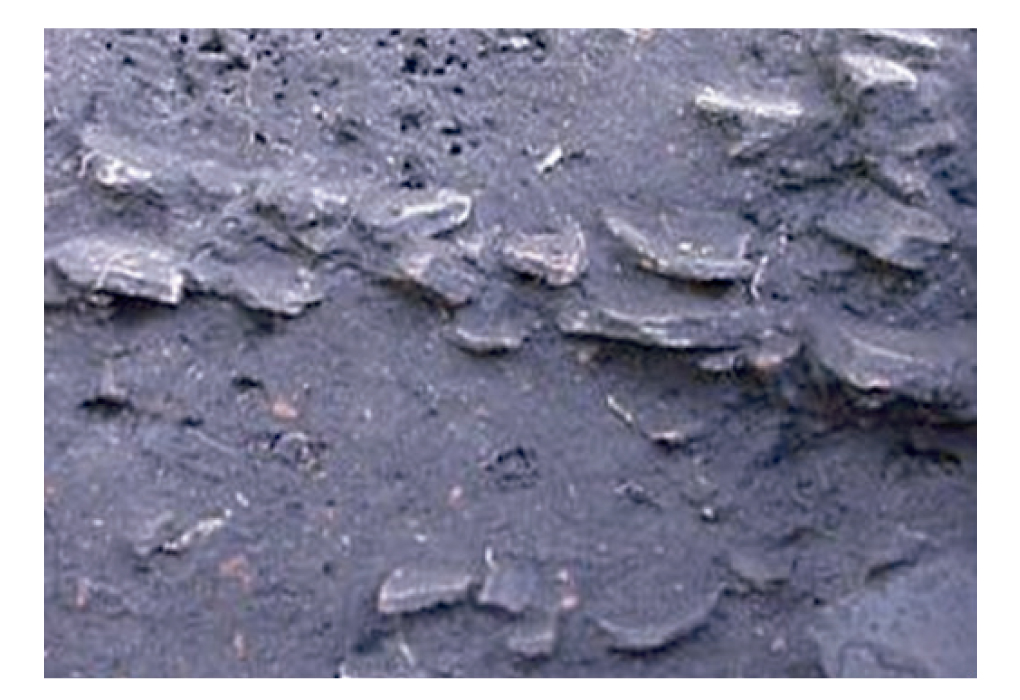}};
    \draw[black, thick] (0,0) circle (\r);
  \end{tikzpicture}
};

\node[below=0.0cm of A] {\footnotesize Earthworks};
\node[below=0.0cm of B] {\footnotesize ADEs};

\node (Y1) at (6,2.5) {%
  \begin{tikzpicture}
    \clip (0,0) circle (\r);
    \node at (0,0) {\includegraphics[width=\d,height=\d]{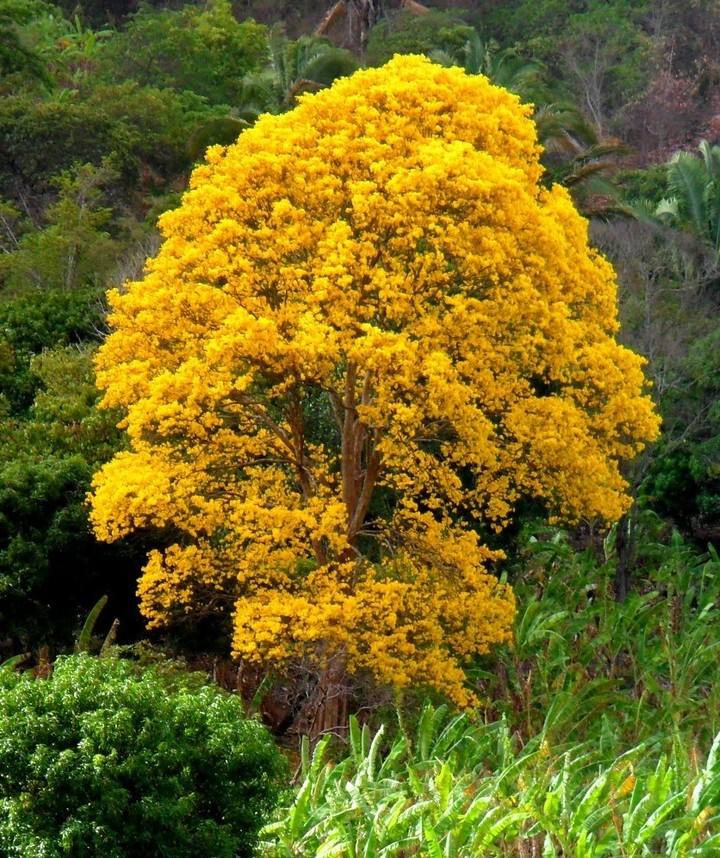}};
    \draw[black, thick] (0,0) circle (\r);
  \end{tikzpicture}
};

\node (Y2) at (6,0) {%
  \begin{tikzpicture}
    \clip (0,0) circle (\r);
    \node at (0,0) {\includegraphics[width=\d,height=\d]{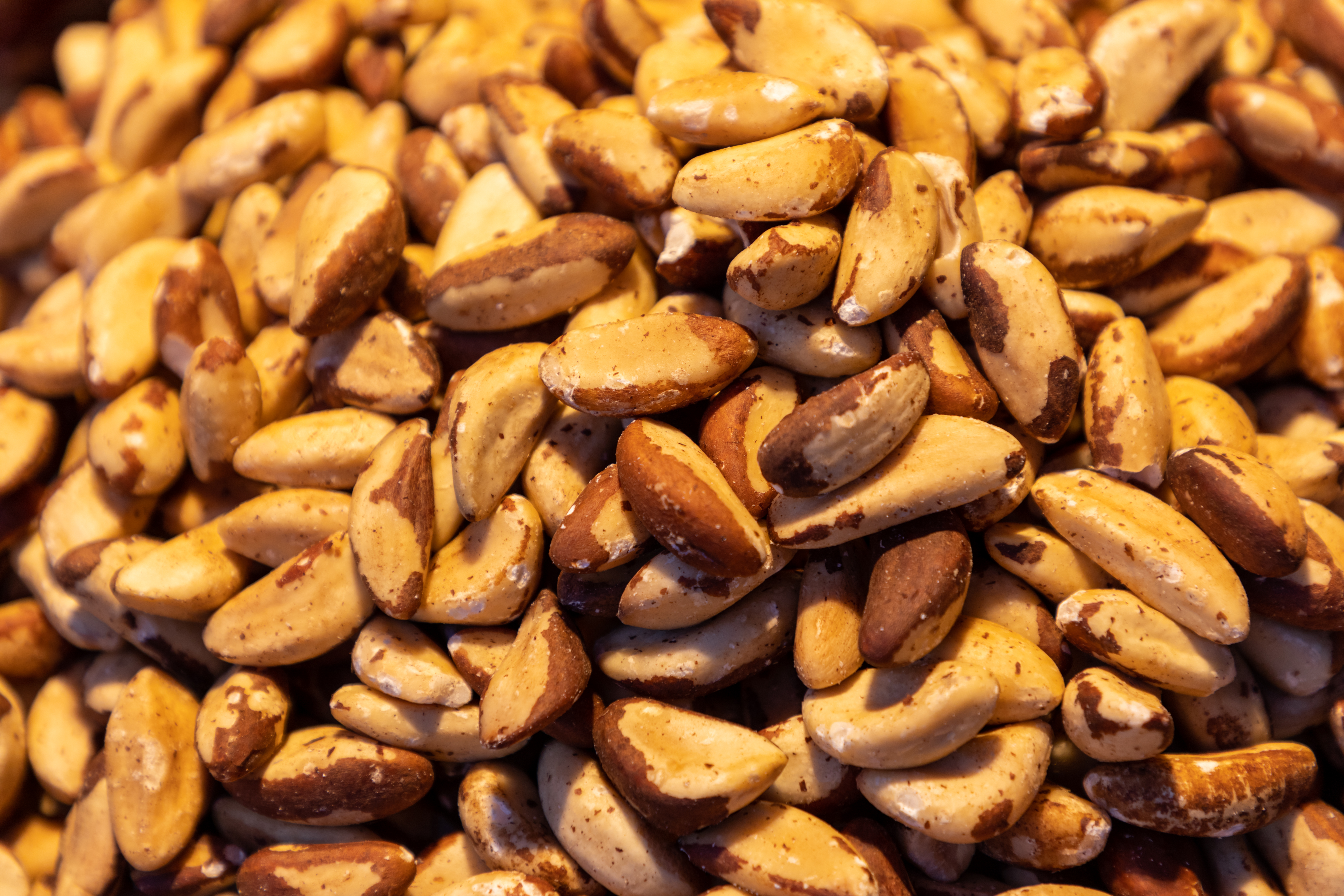}};
    \draw[black, thick] (0,0) circle (\r);
  \end{tikzpicture}
};

\node (Y3) at (6,-2.5) {%
  \begin{tikzpicture}
    \clip (0,0) circle (\r);
    \node at (0,0) {\includegraphics[width=\d,height=\d]{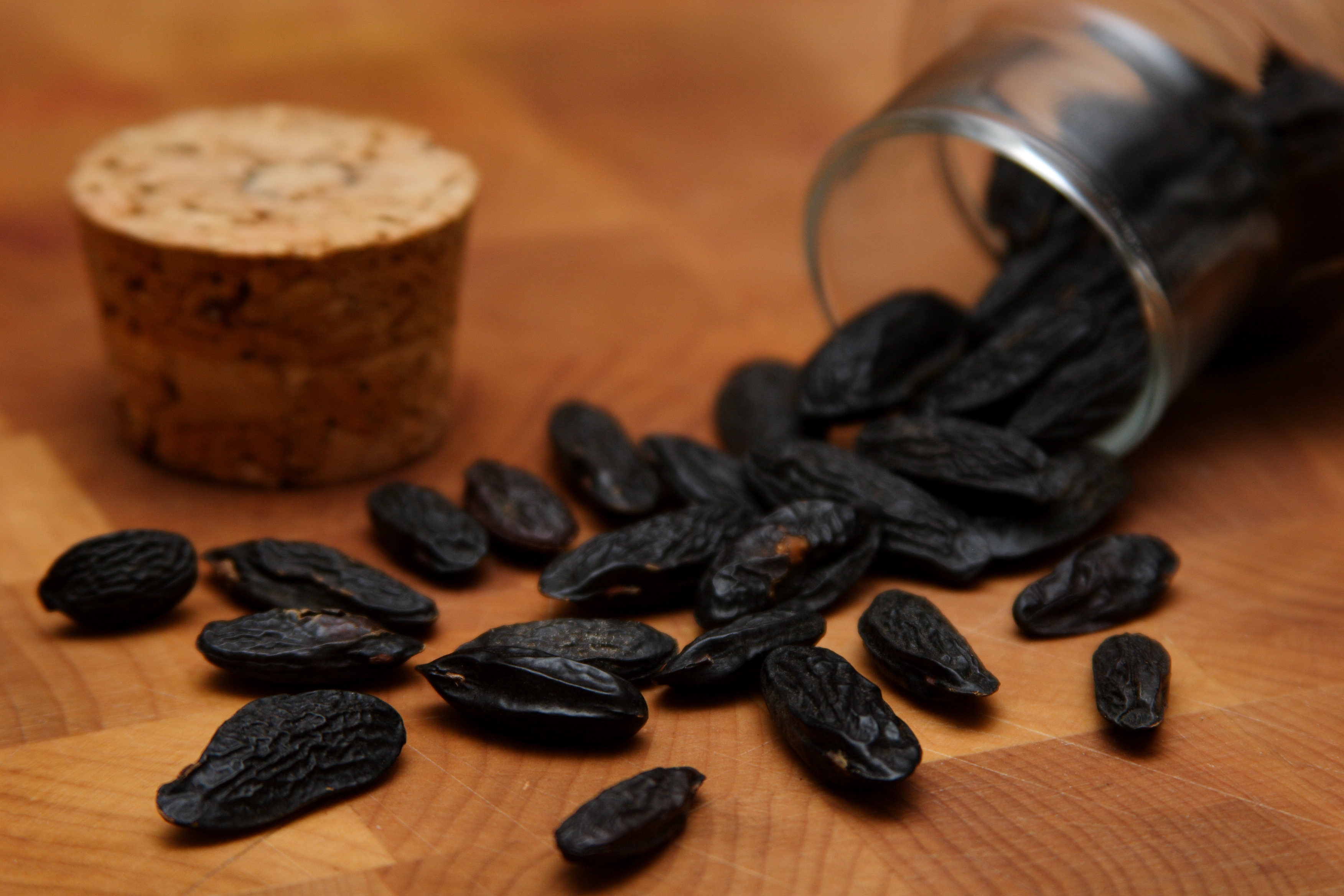}};
    \draw[black, thick] (0,0) circle (\r);
  \end{tikzpicture}
};

\node[below=0.0cm of Y1] {\footnotesize \textit{H. serratifolius}};
\node[below=0.0cm of Y2] {\footnotesize \textit{B. excelsa}};
\node[below=-0.1cm of Y3] {\footnotesize \textit{D. odorata}};

\draw[negedge] (A) -- node[midway, above, sloped, text=red!75!black] {\bm{$\eta_{31}$}} (Y1);
\draw[posedge] (B) -- node[pos=0.42, anchor=west, above, sloped, text=myblue] {\bm{$\eta_{32}$}} (Y1);

\draw[posedge] (A) -- node[midway, above, sloped, text=myblue] {\bm{$\eta_{41}$}} (Y2);
\draw[posedge] (B) -- node[pos=0.38, anchor=west, above, sloped, text=myblue] {\bm{$\eta_{42}$}} (Y2);

\draw[negedge] (A) -- node[midway, above, sloped, text=red!75!black] {\bm{$\eta_{51}$}} (Y3);
\draw[posedge] (B) -- node[midway, above, sloped, text=myblue] {\bm{$\eta_{52}$}} (Y3);

\end{tikzpicture}
}
\caption{}
\label{f:graph-amazon}
\end{subfigure}
\hfill
\begin{subfigure}[c]{0.5\textwidth}
\centering
\includegraphics[width=\textwidth]{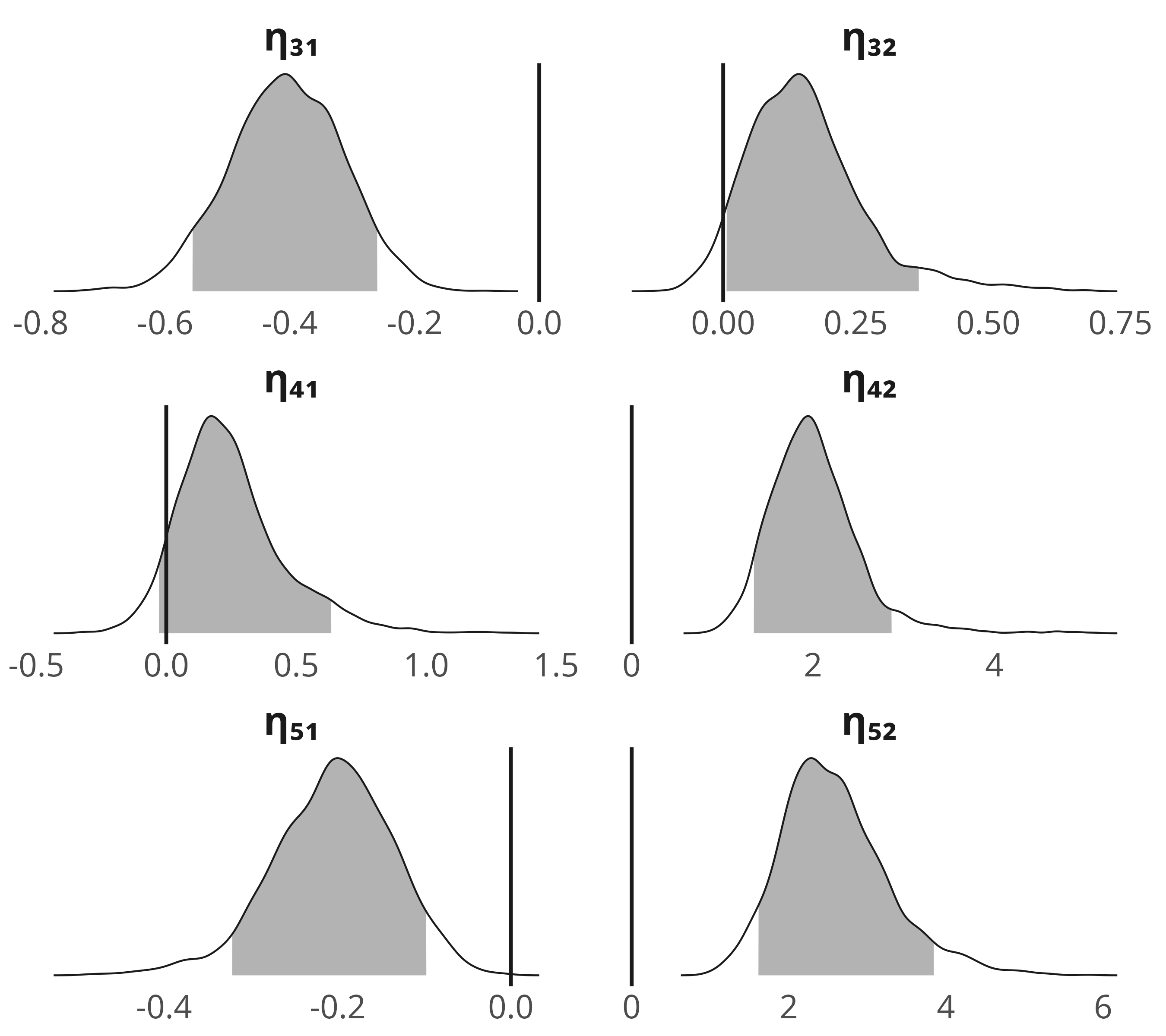}
\caption{}
\label{f:dens_etas}
\end{subfigure}

\caption{
(a) Graph representing the modeled dependency structure between archaeological sites (ADEs and earthworks) and species occurrences. Each edge is labeled with its corresponding parameter $\eta$. Blue edges indicate positive estimated parameters and red edges indicate negative estimated parameters. (b) Posterior densities of the parameters $\eta$. The shaded regions represent the equal tail probabilities 90\% credibility intervals, while the solid vertical lines mark zero. Rows correspond, respectively, to \textit{H. serratifolius}, \textit{B. excelsa}, and \textit{D. odorata}. The first column shows the $\eta$ effects associated with earthworks, whereas the second column corresponds to ADEs. Image credits: earthworks photograph by Sanna Saunaluoma (CC BY-SA 3.0); ADE photograph by Bruno Glaser (CC BY 4.0); \textit{H. serratifolius} photograph by Macelo Costa (CC BY 4.0); \textit{B. excelsa} nuts photograph by Dietmar Rabich (CC BY-SA 4.0); and \textit{D. odorata} seeds photograph by Mecredis / Fred Benenson (CC BY-SA 3.0). Images obtained from Wikimedia Commons and iNaturalist and cropped by the authors.
}
\label{fig:graph-densities}
\end{figure}

\begin{table}[htbp]
\centering
\caption{Posterior summary statistics for the model parameters. CI denotes 90\% credibility intervals.}
\centering
\fontsize{9}{11}\selectfont
\renewcommand{\arraystretch}{0.88}
\begin{tabular}[t]{clcccc}
\toprule
Parameter & Covariate & Mean & Sd. Deviation & Median & CI\\
\midrule
\addlinespace[0.2em]
\multicolumn{6}{l}{\textbf{Earthwork}}\\
\hspace{1em}$\beta_0$ & Intercept & -4.76 & 0.15 & -4.75 & (-5.03 ; -4.52)\\
\hspace{1em}$\beta_1$ & Precip.Dri.Qua. & -0.97 & 0.13 & -0.96 & (-1.19 ; -0.77)\\
\hspace{1em}$\beta_2$ & CationConc. & 1.75 & 0.06 & 1.75 & (1.66 ; 1.85)\\
\hspace{1em}$\delta_0$ & Intercept & -5.80 & 0.31 & -5.80 & (-6.30 ; -5.31)\\
\hspace{1em}$\delta_1$ & TreeCover & -1.53 & 0.11 & -1.52 & (-1.72 ; -1.36)\\
\hspace{1em}$\delta_2$ & Dist.Roads & -4.45 & 0.38 & -4.44 & (-5.07 ; -3.83)\\
\hspace{1em}$\lambda^{*}$ & - & 19,290.85 & 2,983.96 & 19,047.02 & (14,912.79 ; 24,109.78)\\
\addlinespace[0.2em]
\multicolumn{6}{l}{\textbf{ADE}}\\
\hspace{1em}$\beta_0$ & Intercept & -4.31 & 0.27 & -4.34 & (-4.70 ; -3.86)\\
\hspace{1em}$\beta_1$ & RiverDistance & -0.69 & 0.08 & -0.69 & (-0.82 ; -0.56)\\
\hspace{1em}$\beta_2$ & Elevation & -1.99 & 0.28 & -1.99 & (-2.45 ; -1.52)\\
\hspace{1em}$\delta_0$ & Intercept & 2.16 & 0.77 & 2.10 & (0.98 ; 3.52)\\
\hspace{1em}$\delta_1$ & TreeCover & -5.09 & 1.38 & -4.99 & (-7.51 ; -3.00)\\
\hspace{1em}$\delta_2$ & Dist.Roads & -0.81 & 0.14 & -0.80 & (-1.05 ; -0.59)\\
\hspace{1em}$\lambda^{*}$ & - & 3,763.15 & 824.90 & 3,830.19 & (2,469.58 ; 5,060.38)\\
\addlinespace[0.2em]
\multicolumn{6}{l}{\textbf{\textit{H. serratifolius}}}\\
\hspace{1em}$\beta_0$ & Intercept & -1.33 & 0.47 & -1.35 & (-2.08 ; -0.52)\\
\hspace{1em}$\beta_1$ & Precip.Wet.Month & -0.17 & 0.08 & -0.17 & (-0.31 ; -0.04)\\
\hspace{1em}$\beta_2$ & SoilpH & -0.24 & 0.06 & -0.23 & (-0.35 ; -0.15)\\
\hspace{1em}$\eta_1$ & ProximityEarth & -0.41 & 0.09 & -0.40 & (-0.56 ; -0.26)\\
\hspace{1em}$\eta_2$ & ProximityADE & 0.15 & 0.11 & 0.14 & (0.00 ; 0.37)\\
\hspace{1em}$\delta_0$ & Intercept & -3.23 & 0.64 & -3.12 & (-4.33 ; -2.30)\\
\hspace{1em}$\delta_1$ & TreeCover & -0.28 & 0.08 & -0.28 & (-0.41 ; -0.16)\\
\hspace{1em}$\delta_2$ & Dist.Roads & -1.63 & 0.19 & -1.62 & (-1.94 ; -1.32)\\
\hspace{1em}$\lambda^{*}$ & - & 2,826.51 & 1,331.10 & 2,587.70 & (1,039.84 ; 5,690.95)\\
\addlinespace[0.2em]
\multicolumn{6}{l}{\textbf{\textit{B. excelsa}}}\\
\hspace{1em}$\beta_0$ & Intercept & -1.98 & 0.25 & -1.96 & (-2.39 ; -1.59)\\
\hspace{1em}$\beta_1$ & Elevation & -1.93 & 0.31 & -1.92 & (-2.45 ; -1.45)\\
\hspace{1em}$\beta_2$ & CoarseFragments & -0.27 & 0.10 & -0.27 & (-0.44 ; -0.11)\\
\hspace{1em}$\eta_1$ & ProximityEarth & 0.25 & 0.21 & 0.22 & (-0.03 ; 0.64)\\
\hspace{1em}$\eta_2$ & ProximityADE & 2.01 & 0.50 & 1.96 & (1.35 ; 2.87)\\
\hspace{1em}$\delta_0$ & Intercept & -4.06 & 0.42 & -4.04 & (-4.78 ; -3.41)\\
\hspace{1em}$\delta_1$ & TreeCover & -0.22 & 0.06 & -0.22 & (-0.32 ; -0.13)\\
\hspace{1em}$\delta_2$ & Dist.Roads & -2.59 & 0.25 & -2.59 & (-2.99 ; -2.19)\\
\hspace{1em}$\lambda^{*}$ & - & 2,868.94 & 1,011.93 & 2,684.04 & (1,538.81 ; 4,871.24)\\
\addlinespace[0.2em]
\multicolumn{6}{l}{\textbf{\textit{D. odorata}}}\\
\hspace{1em}$\beta_0$ & Intercept & -1.32 & 0.30 & -1.32 & (-1.81 ; -0.83)\\
\hspace{1em}$\beta_1$ & Precip.Feb. & 0.20 & 0.10 & 0.19 & (0.03 ; 0.36)\\
\hspace{1em}$\beta_2$ & OrganicCarbon & -0.85 & 0.20 & -0.85 & (-1.20 ; -0.52)\\
\hspace{1em}$\eta_1$ & ProximityEarth & -0.20 & 0.07 & -0.20 & (-0.32 ; -0.10)\\
\hspace{1em}$\eta_2$ & ProximityADE & 2.59 & 0.68 & 2.51 & (1.60 ; 3.85)\\
\hspace{1em}$\delta_0$ & Intercept & 1.87 & 1.03 & 1.81 & (0.32 ; 3.74)\\
\hspace{1em}$\delta_1$ & TreeCover & -5.40 & 1.85 & -5.29 & (-8.73 ; -2.60)\\
\hspace{1em}$\delta_2$ & Dist.Roads & -1.07 & 0.20 & -1.06 & (-1.41 ; -0.75)\\
\hspace{1em}$\lambda^{*}$ & - & 261.06 & 55.33 & 254.23 & (182.95 ; 363.53)\\
\bottomrule
\end{tabular}
\label{tab:posterior}
\end{table}

Table~\ref{tab:posterior} summarizes the posterior estimates obtained. Distance to roads emerged as the dominant observability covariate for earthworks, \textit{H. serratifolius}, and \textit{B. excelsa}, with negative effects indicating higher detection near roads. This pattern is consistent with results reported by \citet{peripato2023more}. In contrast, vegetation cover is the main observability factor for \textit{D. odorata} and ADE sites, with higher detection in areas of lower tree cover.

The estimated effects of intensity covariates are broadly consistent with previous findings. Earthworks are negatively associated with precipitation of the driest quarter and positively associated with soil cation concentration \citep{deSouza2018,peripato2023more}. ADE sites show negative associations with river distance and elevation, indicating higher occurrence near rivers and at lower altitudes \citep{Walker2023}. 

Figure~\ref{fig:graph-densities}(b) presents the posterior densities of the parameters \(\eta\), indicating different patterns across species. For \textit{H. serratifolius}, a negative effect is observed for proximity to earthworks, while a small positive effect is associated with ADE sites. For \textit{B. excelsa}, no clear effect is observed for proximity to earthworks, as the 90\% credibility interval for the corresponding parameter includes zero, although the posterior distribution places considerable mass on positive values. In contrast, a strong positive association is observed with proximity to ADE sites. For \textit{D. odorata}, the results indicate a negative association with proximity to earthworks and a positive association with proximity to ADE sites.

Figure~\ref{fig:graph-densities} summarizes these results, showing that the associations with ADEs are consistently positive and generally stronger than those observed for earthworks. These findings are consistent with the high fertility of ADE soils, which are commonly associated with agriculture and long-term sedentary occupations. In contrast, the estimated effects of earthworks were negative for \textit{H. serratifolius} and \textit{D. odorata}. For \textit{H. serratifolius}, a similar pattern was reported by \citet{McMichael2025}, who found a negative correlation between its abundance and predictions of pre-Columbian human influence, although the associated p-value was not sufficient to reject the null hypothesis of zero correlation. ADEs are widely distributed across the Amazon, overlapping more broadly with the distribution of domesticated and useful tree species sampled throughout the basin. This broader spatial correspondence may increase the strength of the detected relationships between ADEs and these species. In contrast, earthworks show a more regionally concentrated distribution, particularly along the southern Amazonian border and in the Brazilian state of Acre, which may partially explain their weaker associations with domesticated species at the Amazon-wide scale \citep{silverman2008,levis2017}.

\begin{figure}[htbp]
  \centerline{\includegraphics[width=0.8\textwidth]{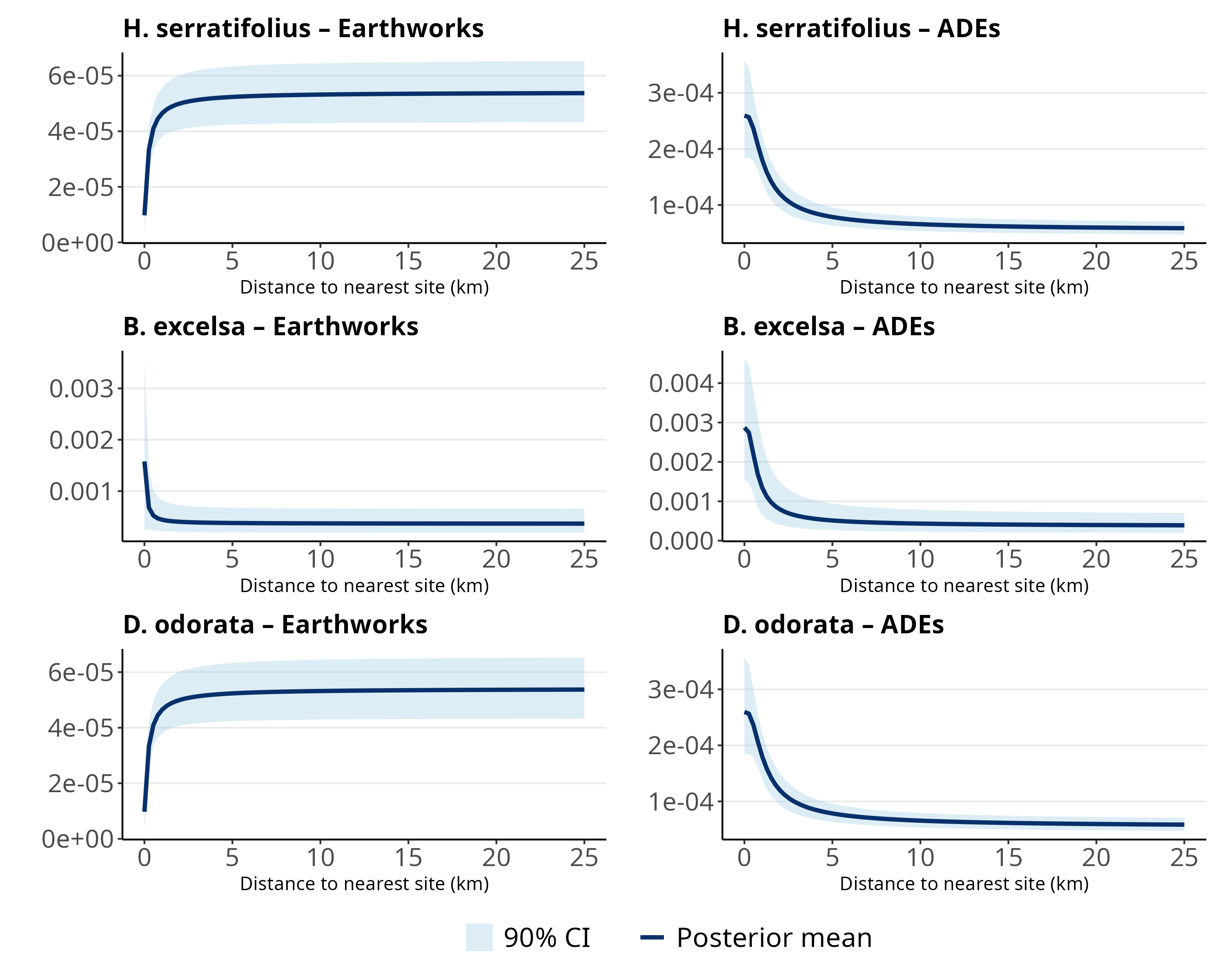}}
\caption{Estimated response curves relating the presence probability in a 1\,km$^2$ cell to the distance from the nearest archaeological site, for each species and type of archaeological site (earthworks and ADEs). The x-axis represents the distance (in km) to the nearest corresponding site. The curves are derived from the fitted model and the estimated Poisson process intensities. The solid line represents the posterior mean, and the shaded region corresponds to the 90\% credibility interval.}
\label{f:response_curve}
\end{figure}

Figure~\ref{f:response_curve} illustrates the estimated relationships between species occurrence probabilities and the distance to the nearest archaeological site. For ease of interpretation, the response curves are shown on the distance scale, obtained by transforming the proximity term $d$ back to the corresponding distances. For instance, for \textit{H. serratifolius} proximity to earthworks is associated with lower estimated presence probabilities. Holding all other covariates fixed, the posterior mean probability of presence in a 1 km\(^2\) cell decreases by 50\% when moving from locations farther than 25 km from the nearest earthwork site to locations approximately 0.4 km away. For \textit{B. excelsa}, the estimated presence probability increases as the proximity to ADEs increases. Keeping all other covariates fixed, the probability of presence increases from approximately 0.00036 for locations farther than 25 km from the nearest ADE site to approximately 0.00136 for locations situated 1 km away, corresponding to an increase of approximately 3.7 times.

An important feature of the proposed model is its ability to estimate both the number and the spatial distribution of unobserved locations. This follows directly from the data augmentation strategy, in which the latent processes \(X_i'\), for \(i = 1,\dots,5\), explicitly represent unobserved occurrences. The probability of unobserved occurrences is computed, for each spatial cell, as the proportion of MCMC iterations in which points from \(X_i'\) are sampled within that cell.

\begin{figure}[htbp]
  \centerline{\includegraphics[width=1\textwidth]{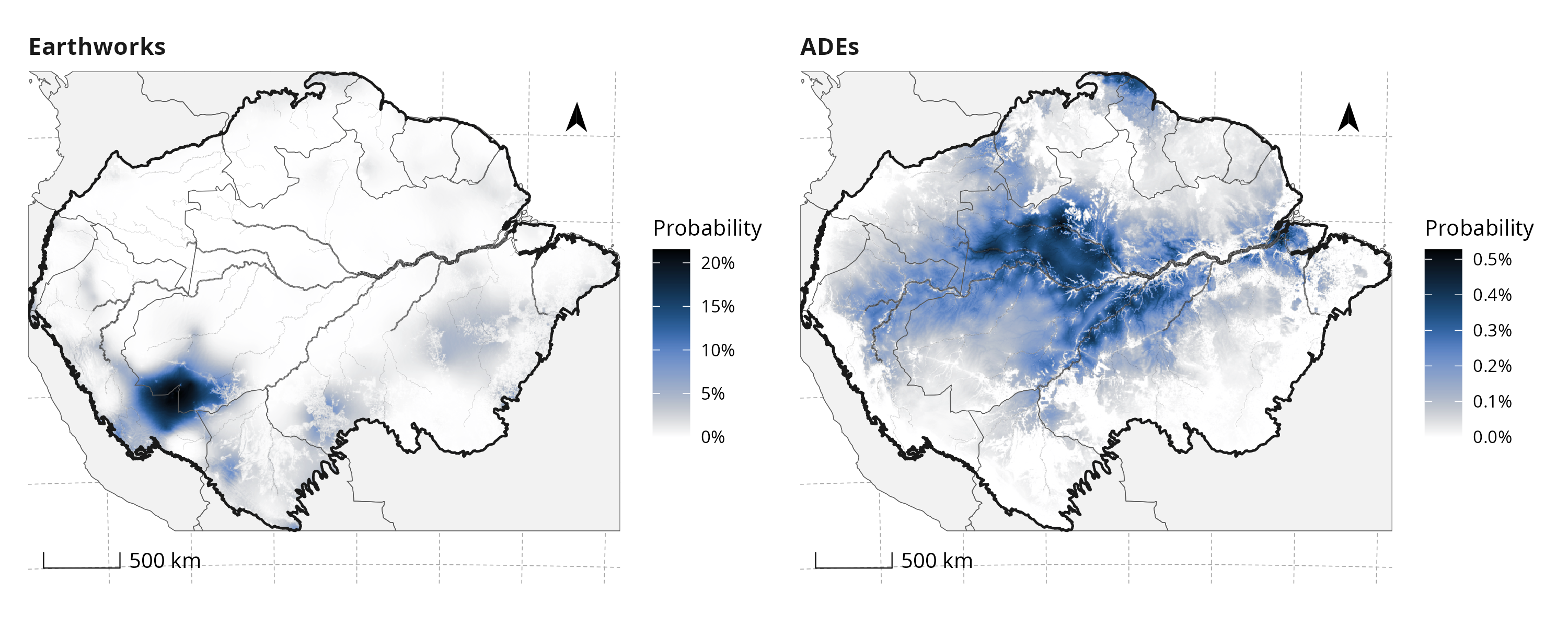}}
  \caption{Posterior predictive probability of occurrence of the latent processes $X_1'$ and $X_2'$ (earthworks and ADEs). Colors represent the proportion of MCMC iterations in which an unobserved occurrence was sampled in each 1\,km$^2$ cell, interpreted as presence probability. Major rivers are shown as gray lines.}
\label{f:pred_heatmap}
\end{figure}

The resulting probability maps for earthworks and ADE sites are presented in Figure~\ref{f:pred_heatmap}. The lower predicted probabilities for ADE sites reflect the posterior estimates of their corresponding \(\lambda^\ast\), whose posterior mean is 3,763.15, compared to 19,290.85 for earthworks. As a consequence, higher probabilities of unobserved occurrences, as well as a larger total number of unobserved points, are expected for earthworks.

In addition, the model allows estimation of the total number of unobserved occurrences for each process. The 90\% credibility interval for \textit{H. serratifolius} ranges from 1,329 to 8,829, while for \textit{B. excelsa} it ranges from 2,823 to 9,889. In contrast, \textit{D. odorata} exhibits a substantially smaller interval, between 219 and 349 unobserved occurrences. For the archaeological sites, the estimated number of unobserved earthworks locations ranges from 4,971 to 6,886, whereas the corresponding interval for ADEs ranges from 340 to 518 occurrences. 

\section{Final remarks}
\label{s:final}

This paper proposes a Bayesian methodology for the joint modeling of multiple spatial point processes under a Bayesian network structure. The proposed approach allows dependence structures between processes to be explicitly represented through model parameters. The proposed model also accommodates presence-only data, where preferential sampling and unobserved occurrences pose additional inferential challenges.

Latent processes are introduced into the model both to account for unobserved occurrences and to obtain an exact inferential scheme without relying on likelihood approximations. This formulation further allows point processes to act as covariates for other processes through both their observed and latent (unobserved) occurrences.

The main contribution of this work is therefore not only to enable joint modeling of multiple point patterns, but to provide a methodology in which relationships can be directly interpreted through estimated parameters. This distinguishes the proposed approach from existing methods that incorporate multiple datasets primarily to improve predictive performance. At the same time, previous studies have shown that exact inference can lead to improved predictive performance compared to models that rely on likelihood approximations \citep{moreira2022analysis,SilvaGamerman2026}.

An additional aspect discussed in this work concerns graph structure learning. Although inference is performed conditionally on a fixed graph \(\mathcal{G}\), posterior inference on parameters \(\eta\) provides a simple mechanism to identify unsupported edges and prune overly connected graphs.

The model was evaluated through simulation studies and an application involving archaeological sites and tree species occurrences in Amazonia. The estimated dependence patterns between processes were broadly consistent with previous findings in the literature, while the latent process formulation additionally allowed inference on unobserved occurrences through probability maps and estimates of the total number of unobserved occurrences. 
Together, these results illustrate the applicability of the proposed methodology and provide evidence of its practical relevance.

This work also gives rise to several directions for future research. From a computational perspective, strategies to improve efficiency could be explored, including parallel implementations of the MCMC scheme in which conditionally independent nodes are sampled simultaneously within each iteration of the algorithm \citep{sabek2019flash}. Another possible extension is the incorporation of undirected edges, relaxing the directional assumption adopted in this work. Finally, introducing spatial autocorrelation components into the point process specification may provide additional flexibility, particularly in settings where important spatial variation is not fully captured by the available covariates.

\section*{Acknowledgements}

The authors thank the Department of Statistics at the Federal University of Minas Gerais (UFMG), where this work was developed. The authors thank professors Marcos Prates and Lilia da Costa for valuable comments throughout the development of this work. They also thank Manolo Quintilhan and Nelson Barrios for discussions that contributed to the selection of species considered in the application. The authors acknowledge Guido Moreira for previous methodological work that served as a foundation for this research. They also thank the National Institute for Space Research (INPE) and the Brazil Data Cube (BDC) project for providing access to the BDC-Lab computational environment used in this work. This paper is based on the master's thesis of the first author, developed under the supervision of the second author and co-supervision of the last author. The first author acknowledges financial support from FAPEMIG, Brazil. The third and last authors acknowledge financial support from the CARBS 2.0 Project (Convergence to Accelerate Research on Biological Sequestration), coordinated by professor Lucas Silva and funded by the University of Oregon through the Amazon Sustainability Center and administered by Finatec, Brazil. The first and second authors also acknowledge financial support from CNPq, Brazil, through projects 302929/2022-8 and INCT 406913/2022-6.

\appendix 
\section*{Appendix}
\addcontentsline{toc}{section}{Appendix} 
\section{Sampling of $\zeta_i$ and $\delta_i$} \label{appendix:coefs}

This section summarizes the sampling schemes adopted for the regression coefficients under the logit and probit links. The resulting conditional distributions follow directly from the augmentation strategies proposed by \citet{polson2012} and \citet{albert1993}.

\subparagraph{Logit link.}

Under the logit link, the augmented formulation can be expressed as a set of logistic regressions and sampling is performed using the Pólya--Gamma augmentation scheme of \citet{polson2012}.

To formalize the binary representation, indicator variables \(y_{ij}\) are introduced, taking value 1 for points classified as successes and 0 for points classified as failures. For the regression associated with \(\zeta_i\), observations in \(x_i \cup x_i'\) correspond to \(y_{ij}=1\), whereas observations in \(u_i\) correspond to \(y_{ij}=0\). For the regression associated with \(\delta_i\), observations in \(x_i\) correspond to \(y_{ij}=1\), while observations in \(x_i'\) correspond to \(y_{ij}=0\).

For the regression associated with \(\zeta_i\), let \(j=1,\dots,n_i\), where \(n_i = n_{x_i} + n_{x_i'} + n_{u_i}\), whereas for \(\delta_i\), let \(j=1,\dots,n_{\tilde{x}i}\), with \(n_{\tilde{x}_i} = n_{x_i} + n_{x_i'}\). For each process \(i=1,\dots,N\), latent Pólya--Gamma variables are introduced according to
\begin{equation}
\omega_j\mid\zeta_i \sim \text{Pólya-Gamma}\!\left(1,\tilde{Z}_i(s)^{\top}\zeta_i\right),
\end{equation}
and
\begin{equation}
\omega_j\mid\delta_i \sim \text{Pólya-Gamma}\!\left(1,W_i(s)^{\top}\delta_i\right).
\end{equation}
where \(\tilde{Z}_i(s)\) denotes the concatenation of the covariate vectors \(Z_i(s)\) and \(d_i(s)\).

Conditional on the augmented variables, both \(\zeta_i\) and \(\delta_i\) have multivariate Normal full conditional distributions,
\begin{equation}
\zeta_i\mid y_i,\omega \sim \mathcal{N}_k\left(m_{\zeta_i},V_{\zeta_i}\right),
\end{equation}
and
\begin{equation}
\delta_i\mid y_i,\omega \sim \mathcal{N}_m\!\left(m_{\delta_i},V_{\delta_i}\right),
\end{equation}
with mean and covariance matrices given by
\begin{equation}
V_{\zeta_i} = \left(\tilde{Z}_i^{\top}\Omega\tilde{Z}_i + B_i^{-1}\right)^{-1},
\end{equation}
\begin{equation}
m_{\zeta_i} = V_{\zeta_i}\left(\tilde{Z}_i^{\top}\kappa_i + B_i^{-1}b_i\right),
\end{equation}
\begin{equation}
V_{\delta_i} = \left(W_i^{\top}\Omega W_i + F_i^{-1}\right)^{-1},
\end{equation}
\begin{equation}
m_{\delta_i} = V_{\delta_i}\left(W_i^{\top}\kappa_i + F_i^{-1}f_i\right).
\end{equation}

Here, \(\Omega\) denotes the diagonal matrix formed by the latent variables \(\omega_j\) and \(\kappa_i = \left(y_{i1}-\frac{1}{2},\dots,y_{in_i}-\frac{1}{2}\right)\). Analogously, \(\kappa_i = \left(y_{i1}-\frac{1}{2},\dots,y_{in_{\tilde{x}_i}}-\frac{1}{2}\right)\) for the observability coefficients. The prior distributions are given by \(\pi(\zeta_i)\sim \mathcal{N}_k(b_i,B_i)\) and \(\pi(\delta_i)\sim \mathcal{N}_m(f_i,F_i)\), where \(b_i\) and \(f_i\) denote prior mean vectors, and \(B_i\) and \(F_i\) denote prior covariance matrices.

\subparagraph{Probit link.}

Under the probit link, sampling follows the augmented Gaussian construction of \citet{albert1993}. The same binary representations described above are adopted, but inference is performed through latent truncated Normal variables.

For \(\zeta_i = (\beta_i,\eta_i)\), latent variables are introduced as
\begin{equation}
\psi_{j} \mid y_i, \zeta_i \sim \mathcal{N}\!\left(\tilde{Z}_i(s)^{\top}\zeta_i\,,1\right),
\end{equation}
whereas for \(\delta_i\),
\begin{equation}
\psi_{j} \mid y_i,\delta_i \sim \mathcal{N}\!\left(W_i(s)^{\top}\delta_i,\,1\right).
\end{equation}

The latent variables are truncated according to the observed binary outcomes, such that \(\psi_j > 0\) if \(y_{ij}=1\), and \(\psi_j \le 0\) if \(y_{ij}=0\). Conditional on the augmented variables, both coefficient vectors again have multivariate Normal full conditional distributions,
\begin{equation}
\zeta_i \mid y_i,\psi \sim \mathcal{N}_k\!\left(\tilde{b}_i,\tilde{B}_i\right),
\end{equation}
and
\begin{equation}
\delta_i \mid y_i,\psi \sim \mathcal{N}_m\!\left(\tilde{f}_i,\tilde{F}_i\right),
\end{equation}
where \(\tilde{B}_i = \left(B_i^{-1} + \tilde{Z}_i^{\top}\tilde{Z}_i\right)^{-1}\), \(\tilde{b}_i =
\tilde{B}_i\left(B_i^{-1}b_i + \tilde{Z}_i^{\top}\psi\right)\), \(\tilde{F}_i = \left(F_i^{-1} + W_i^{\top}W_i\right)^{-1}\), and \(\tilde{f}_i = \tilde{F}_i\left(F_i^{-1}f_i + W_i^{\top}\psi\right)\).

\bibliographystyle{unsrtnat}
\bibliography{references}  

\end{document}